%% file: main.tex
\documentclass[]{article}
\usepackage{graphicx}
\usepackage{hyperref}
\hypersetup{
    colorlinks = true,
    urlcolor   = blue,
    citecolor  = blue,
}
\usepackage{color}
\usepackage[hang,small,center]{caption}
\usepackage{amsbsy}
\usepackage{amssymb}
\usepackage{natbib}
\usepackage{amsmath}

\newcommand{\RomanNumeralCaps}[1]
\linenumbers

\usepackage{multirow}
\usepackage{array}
\usepackage{bbold}
\usepackage{calc}
\usepackage{subcaption}
\usepackage{authblk}
\usepackage{geometry}
\input{commands}

\date{}
\title{Two-scale modelling of two-phase flows based on the Stationary Action Principle and a Geometric Method Of Moments}

  \author[1]{Arthur Loison}
  \author[1]{Teddy Pichard}
  \author[2]{Samuel Kokh}
  \author[2]{Marc Massot}

  \affil[1]{CMAP, CNRS, École polytechnique, Institut Polytechnique de Paris, Palaiseau 91120, France}
  \affil[2]{Université Paris-Saclay, CEA, Service de Génie Logiciel pour la Simulation, 91191, Gif-sur-Yvette, France.}

\begin{document}
\maketitle

\input{0_abstract}



\input{0_intro}

\input{1_two_regimes}
\input{2_polydispersion}

\input{3_oscillation}

\input{4_geometry_dyn}
\input{5_conclusion}

\appendix

\input{SAP_comp_incomp}
\input{SAP_two_scale}

\input{geomom_reconstruction}

\input{SAP_poly}

\input{diff_geom_S_SH}

\input{poly_quad}

\input{SAP_poly_osc}

\input{monodisperse}

\bibliographystyle{abbrvnat}
\bibliography{biblio}

\end{document}

%% file: commands.tex
\newcommand{\vel}{\boldsymbol{u}}
\newcommand{\lag}{\mathcal{L}}
\newcommand{\tEntropy}{\mathcal{H}}

\newcommand{\SigG}{\Sigma\left<G\right>}
\newcommand{\SigH}{\Sigma\left<H\right>}
\newcommand{\SigPerpEq}{\Sigma_{\perp,0}}
\newcommand{\DSigPerp}{(\Delta\Sigma_{\perp})}
\newcommand{\DtSigPerp}{(\Delta_t\Sigma_{\perp})}
\newcommand{\SigHPerpEq}{(\Sigma\left<H\right>)_{\perp,0}}
\newcommand{\DSigHPerp}{(\Delta\Sigma\left<H\right>)_{\perp}}
\newcommand{\DtSigHPerp}{(\Delta_t\Sigma\left<H\right>)_{\perp}}
\newcommand{\VFracd}{\alpha_1^d}

\newcommand{\VarTraj}{\boldsymbol{\eta}}
\newcommand{\Action}{\mathcal{A}}

\newcommand{\bcdot}{\boldsymbol{\cdot}}
\newcommand{\bnabla}{\boldsymbol{\nabla}}

%% file: 0_abstract.tex
\begin{abstract}
In this contribution, we introduce a versatile formalism to derive unified two-phase models describing both the separated and disperse regimes. It relies on the stationary action principle and interface geometric variables.
The main ideas are introduced on a simplified case where all the scales and phases have the same velocity and that does not take into account large-scale capillary forces.
The derivation tools yield a proper mathematical framework through hyperbolicity and signed entropy evolution.
The formalism encompasses a hierarchy of small-scale reduced-order models based on a statistical description at a mesoscopic kinetic level and is naturally able to include the description of a disperse phase with polydispersity in size. 
This hierarchy includes both a cloud of spherical droplets and non-spherical droplets experiencing a dynamical behaviour through incompressible oscillations. 
The associated small-scale variables are moments of a number density function resulting from the Geometric Method Of Moments (GeoMOM).
This method selects moments as small-scale geometric variables compatible with the structure and dynamics of the interface; they are defined independently of the flow topology and, therefore, this model pursues the goal of unifying the modelling of a fully-coupled two-scale flow.
It is particularly showed that the resulting dynamics provides closures for the interface area density equation obtained from the averaging approach.
The extension to mass transfer from one scale to the other including capillary phenomena, as well as the extension to multiple velocities are possible and proposed in complementary works.
\end{abstract}

%% file: 0_intro.tex
\section{Introduction}
%
%
%
Two-phase flows encompass a wide range of physical phenomena, presenting dynamics involving a large spectrum of scales~\citep{scardovelli_direct_1999,tomar_multiscale_2010,shinjo_simulation_2010,dumouchel_multi-scale_2015,cordesse_validation_2020,sakano_evaluation_2022,estivalezes_phase_2022}.
We particularly identify two different regimes based on the topology/geometry of the fluid interfaces~\citep{ishii_thermo-fluid_1975}: separated-phases flows, where the material interface is described at the bulk fluid scale (hereafter called large scale), and disperse flows, characterized by the presence of disperse phases within a carrier fluid, resulting in interfaces with significantly smaller scales (hereafter called small scale). 

Simulating such flows is crucial for many industrial processes such as combustion chambers where liquid fuel is injected at high velocity and pressure forming a cloud of liquid droplets further away from the nozzle or nucleate boiling in nuclear reactors~\citep{jamet_towards_2010, providakis_characterization_2012, fiorina_modeling_2016}.
Unfortunately, Direct Numerical Simulation (DNS) capturing the smallest interfacial variations is currently limited to academic configurations and impractical in industrial contexts. We address these challenges through the development of new modelling approaches.
%
One such approach, known as sub-scale modelling, consists in modelling the sub-scale phenomena directly at the bulk scale. Various approaches have been proposed for both separated and disperse flow regimes, such as two-fluid systems of equations~\cite{baer_two-phase_1986,raviart_non-conservative_1995,saurel_modelling_2017}. But, they often offer limited information about the interfacial structures and are only adapted to one of the two regimes. 
Several works aim at coupling both regimes~\citep{devassy_atomization_2015,le_touze_compressible_2020} and sometimes include other modelling approaches such as Lagrangian models \citep{lebas_numerical_2009}. In the mixed regime zone, when the separated regime transitions to the disperse regime, the coupling processes involve very different descriptions, variables as well as a series of coupling parameters. They have to be tuned depending on the configuration of interest, which prevents the possibility of predictive simulations.  The present work aims at providing a unified formalism that naturally degenerates towards separated and disperse flow models, and that can handle the mixed zone where both large and small scales are present.

Our first contribution is to propose a versatile framework to derive unified two-scale two-phase models based on Hamilton's Stationary Action Principle (SAP)~\citep{herivel_derivation_1955, serrin_mathematical_1959, salmon_practical_1983, bedford_hamiltons_1985, fosdick_kinks_1991, gavrilyuk_hyperbolic_1998, gouin_hamiltons_1999, gavrilyuk_mathematical_2002, berdichevsky_variational_2009, drui_small-scale_2019,gouin_introduction_2020,cordesse_diffuse_2020} and the second principle of thermodynamics.
The originality of the approach relies on the models' ability to account for the presence of both separated and disperse regimes at the same time.
This is made possible by considering an artificial mixture in which small-scale and large-scale interfaces coexist within the same physical domain.
The length-scale threshold that separates the large-scale dynamics of the bulk phase from the dynamics of the small-scale depends on the physics and is not discussed in this work.
By employing Hamilton's SAP and an evolution equation for the mathematical entropy of the model, we can obtain a dissipative system that captures the behaviour of this multiphase medium and provides a good mathematical framework to analyse such models.

The second contribution of this work lies in the ability to plug into the framework a hierarchy of small-scale models for the disperse flow regime. 
We notably extend the Geometric Method Of Moments (GeoMOM) of Essadki~\citep{essadki_statistical_2019} which connects geometric description of the small-scale interfaces \citep{pope_evolution_1988,drew_evolution_1990} with the statistics of a spray of droplets \citep{massot_counterflow_1998,laurent_multi-fluid_2001,fox_multiphase_2007,massot_eulerian_2007}.
We particularly propose a reduced-order moment model for the dynamics of the small-scale inclusions that degenerates toward an existing moment models for the disperse regime in the static minimal surface configuration of spherical polydisperse droplets~\citep{kah_high_2015,essadki_high_2018}, but can also handle the dynamic case of incompressible oscillation of droplets at the kinetic level following~\cite{orourke_tab_1987} to model deformed inclusions.
The key ingredient is the proper choice of variables, which are moments of this kinetic description, but can also be identified as surface-averaged geometric quantities related to the interface dynamics independently of the flow topology.
Thus, it paves the way to the modelling of the mixed zone, where the shape of  droplets and ligaments depart significantly from a spherical form.



Lastly, we demonstrate the compatibility of the obtained small-scale models with several existing models from the literature that describe the evolution of the interfacial area density in the flow. It offers a new way on selecting the adequate variables for interface dynamics description and reaches interesting closures, usually out of reach, for the interfacial area density equation.

The definition of the two-scale mixture and the derivation of its dynamics with Hamilton's SAP is presented in Section~\ref{sec:two-scale}.
Then, the small-scale modelling of polydisperse sprays with GeoMOM is proposed in Section~\ref{sec:polydisperse}.
We pursue with an extension of the polydisperse model of spherical droplets into a spray of oscillating droplets in Section~\ref{sec:kinetic-model-oscill}. Finally, the dynamics of interface area density of these models is discussed and compared with models of the literature in Section~\ref{sec:geometry_dynamics}.

%% file: 1_two_regimes.tex
\section{Two-scale two-phase model for both separated and disperse regimes}
\label{sec:two-scale}
We model a mixture composed of large-scale liquid and gaseous phases in a separated regime and a small-scale liquid phase.
We particularly assume that
\begin{align}
    & \text{\textbullet\: the small-scale liquid phase occupies a small volume of the local mixture.
    } \label{hyp:small_vol} \tag{H0}
\end{align}
The regime of the small-scale phase is not prescribed here, but it is eventually described by a disperse regime model.
The main goal of this first modelling task is to propose a framework to account for the presence of both large-scale and small-scale phases at the same location.

\subsection{Two-scale modelling assumptions}
We propose successive sets of assumptions.
%
First, regarding the fluid dynamics:
\begin{align}
    & \text{\textbullet\: all phases have the same mean velocity; 
    } \label{hyp:ls_shared_vel} \tag{H1a} \\
    & \text{\textbullet\: the small-scale velocity variations are small;
    } \label{hyp:low_vel_diff} \tag{H1b} \\
    & \text{\textbullet\: the small-scale liquid phase is incompressible.
    }  \label{hyp:incomp} \tag{H1c}
\end{align}

%
Following hypotheses (\ref{hyp:ls_shared_vel}) and (\ref{hyp:low_vel_diff}), we do account for velocities that naturally appear at small scale (\textit{e.g.} droplets' drag in the disperse regime) for the sake of clarity of the exposition of the new ideas of the paper.
Potential extensions are discussed in the conclusion.
The incompressibility of the small-scale is discussed in appendix \ref{app:incomp} to justify (\ref{hyp:incomp}).
It impacts the physics at two levels.
The first level pertains to sound propagation that is cancelled at the small scale.
It is especially relevant in the disperse regime where no sound can propagate from one inclusion to another as underlined in \cite{saurel_modelling_2017}.
The second level is related to the volume occupation of the disperse phase and the evolution of its volume fraction.

Second, the following hypotheses are formulated solely for the sake of simplicity, but are not limiting for the models developed in the present work: 
\begin{align}
     & \text{\textbullet\: each phase is isothermal with barotropic equations of state;}  \label{hyp:isothermal} \tag{H2a}     \\
     & \text{\textbullet\: large-scale capillarity effects are not modelled;}  \label{hyp:ls_no_capillarity} \tag{H2b}                             \\
     & \text{\textbullet\: no mass exchanges between the phases are taken into account;}  \label{hyp:ls_no_mass} \tag{H2c}
\end{align}

Let us remark that \eqref{hyp:ls_shared_vel} and \eqref{hyp:isothermal} correspond to the thermal and kinematic equilibria as studied in \cite{chanteperdrix_compressible_2002, caro_dinmod_2005, drui_small-scale_2019}.
The effects neglected by \eqref{hyp:isothermal}, \eqref{hyp:ls_no_capillarity} and \eqref{hyp:ls_no_mass} have been studied for instance in \cite{perigaud_compressible_2005, gaillard_interfaces_2015, schmidmayer_model_2017, cordesse_contribution_2020} and \cite{caro_dinmod_2005, pelanti_arbitrary-rate_2022}.
The model can be extended to include these effects, but they are not necessary for the two-scale modelling framework at stake in this work. 
The particular modelling of mass transfers between the large scale and the small scale is also not considered but is eventually discussed in the conclusion.

We denote with the subscript $k=1$ the liquid and $k=2$ the gaseous phase.
An additional superscript $d$ identifies the small-scale liquid phase, where the letter $d$ stands for ``droplets'' or ``disperse'' liquid phase.
Following~\eqref{hyp:isothermal}, each large scale fluid $k=1,2$ is equipped with a barotropic equation of state of the form $\rho_k \mapsto e_k (\rho_k)$, where $\rho_k$ and $e_k$ are the density and the barotropic potential of the phase $k=1,2$.
The pressure $p_k$ of the large scale phase $k=1,2$ is then defined by $p_k = \rho_k^2 e_k'(\rho_k)$.
We assume that $p_k'(\rho_k)>0$ so that the sound speed $c_k$ associated with the phase $k=1,2$ is $c_k=\sqrt{p_k'(\rho_k)}$.
With $\alpha_k$ the volume fraction of the phase $k=1,2,1^d$, we consider a two-scale mixture such that it obeys the volume constraint
\begin{equation}
    \alpha_1 + \alpha_2 + \alpha_1^d = 1,
\end{equation}
and its density $\rho$ is obtained from the effective phase densities $m_k := \alpha_k\rho_k$,
\begin{equation}
    m_1 + m_2 + m_1^d = \rho.
\end{equation}
Remark that the large-scale volume fractions $\alpha_1, \alpha_2$ range between $0$ and $1$ while the small-scale volume fraction $\alpha_1^d$ is confined to small values $\alpha_1^d\ll 1$ as assumed by \eqref{hyp:small_vol}.
We denote $Y_k$ the mass fraction such that the specific barotropic potential $e$ of the two-phase material is defined by
\begin{equation}
  e =  
    Y_1 e_1(\rho_1)
    +
    Y_2 e_2(\rho_2)
    +
    Y_1^d e_1(\rho_1^d).
\end{equation}
Under assumptions (\ref{hyp:ls_shared_vel}) and (\ref{hyp:low_vel_diff}), there is a unique velocity $\vel$ describing the mixture
\begin{equation}
    \label{eq:unique_vel}
    \vel:= \vel_1 = \vel_2 = \vel_1^d.
\end{equation}
Then, (\ref{hyp:ls_no_mass}) provides the mass conservation of each phase
\begin{equation}\label{eq:mass_cons}
    \partial_t m_k + \bnabla \bcdot (m_k \vel) = 0,
    \quad
    \text{for} \quad k=1,2,1^d.
\end{equation}
Summing these equations provides the total mass conservation equation
\begin{equation}
    \label{eq:mass_conservation}
    \partial_t \rho + \bnabla_{\boldsymbol{x}}\bcdot (\rho \vel)=0.
\end{equation}
Denoting $D_t(\cdot)=\partial_t(\cdot) + \vel \bcdot \bnabla (\cdot)$ the material time derivative, the incompressibility of the small-scale liquid phase (\ref{hyp:incomp}) reads
\begin{equation}
    \label{eq:hyp_incom_dtrho1d}
    D_t \rho_1^d = 0.
\end{equation}
Then liquid volume fraction of the small-scale is conserved
\begin{equation}\label{eq:alpha1d_constraint}
    \partial_t \alpha_1^d + \bnabla \bcdot (\alpha_1^d \vel) = 0.
\end{equation}
\subsection{Hamilton's Stationary Action Principle}
We introduce a novel framework for two-scale modelling  in the spirit of \citep{gavrilyuk_new_1999,drui_small-scale_2019} by using Hamilton's SAP.
We set the kinetic and potential energies of the phase $k$ as
\begin{equation}
    \label{eq:two_scale_energies}
    E_k^{kin}:= \frac{1}{2}m_k \vel^2, \quad  E_k^{pot}:= m_k e_k\left(\frac{m_k}{\alpha_k}\right),
\end{equation}
and the mixture kinetic and potential energies are $\sum_k E_k^{kin}$ and $\sum_k E_k^{pot}$.
This enables the definition of the Lagrangian energy $\lag_k=E_k^{kin}-E_k^{pot}$ of the phase $k$ and a mixture Lagrangian energy $\lag=\sum_k\lag_k$.
Capillarity energy is not taken into account as we lack information about the small-scale interface area density.
These effects will be considered later in Section \ref{sec:polydisperse}, where we refine the small-scale model.
Thereby, the Lagrangian associated with our system reads
\begin{equation}
    \label{eq:lagrangian_two_scale}
    \lag
    =
    \lag_1\left(\alpha_1, m_1, \vel\right)
    +\lag_2\left(\alpha_2, m_2, \vel\right)
    +\lag_1^d\left(m_1^d, \rho_1^d, \vel\right).
\end{equation}
The dependency on $\rho_1^d=m_1^d/\alpha_1^d$ was added to take advantage of the constraint $D_t \rho_1^d = 0$ in the derivation with Hamilton's SAP.
Then, the dependencies of the small-scale Lagrangian are
$\lag_1^d=\tfrac{1}{2}m_1^d \vel^2 - m_1^d e_1(\rho_1^d).$

%
%
Hamilton's SAP provides the momentum equation by minimizing the Lagrangian's action, i.e. the integral of the Lagrangian over a space domain $\Omega_{\boldsymbol{x}}$ and a time interval $[0,T]$.
This minimization takes place under the mass conservation~\eqref{eq:mass_cons} and incompressibility constraints~\eqref{eq:alpha1d_constraint}.
It leads to the following two equations (see appendix \ref{app:SAP} for details)
\begin{equation}
    \label{eq:unevaluated_two_scale_system}
    \begin{cases}
        \partial_t\boldsymbol{K}+\bnabla\bcdot(\boldsymbol{K} \otimes \vel) - \bnabla( \lag^*+\lag_2^*+\lag_1^{d*}-\alpha_1^d\partial_{\alpha_2}\lag_2) = 0, \\
        \partial_{\alpha_1}\lag_1-\partial_{\alpha_2}\lag_2 = 0,
    \end{cases}
\end{equation}
where $\boldsymbol{K}=\partial_{\vel} \lag$, $\lag_k^* = m_k(\partial_{m_k} \lag_k)-\lag_k$ and $\lag^*=\sum_k \lag_k^*$.
Remark that the divergence of a matrix $\boldsymbol{A}$ is a vector denoted $\bnabla\bcdot\boldsymbol{A}$ which here evaluates to $\bnabla\bcdot\boldsymbol{A}=(\partial_{x_j}A_{ij})$ with the summation on repeated indexes.
The choice of energies \eqref{eq:two_scale_energies} yields
\begin{equation}
    \begin{aligned}
        \boldsymbol{K}&=\rho \vel,
        &\lag_1^* &= -\alpha_1 p_1,
        &\partial_{\alpha_1}\lag_1 &= p_1,
        \\
        \lag_1^{d,*} &= 0,
        &\lag_2^* &= -\alpha_2 p_2,
        &\partial_{\alpha_2}\lag_2 &= p_2.
    \end{aligned}
\end{equation}
Including the constraints (\ref{eq:mass_cons}-\ref{eq:alpha1d_constraint}) and evaluating the first equation of \eqref{eq:unevaluated_two_scale_system} leads to
\begin{equation}
    \label{eq:two_scale-incomp-cons}
    \begin{cases}
        \setlength{\arraycolsep}{0pt}
        \begin{array}{lll}
            \partial_t  m_k &\:+ \:\bnabla \bcdot  (m_k \vel)&\:=0,
            \qquad
            k=1,2,1^d, \\
            \partial_t  \alpha_1^d &\:+\: \bnabla \bcdot  (\alpha_1^d \vel)& \:=0,\\
            \partial_t  (\rho \vel) &\:+\: \bnabla \bcdot  (\rho \vel \otimes  \vel + p  \boldsymbol{I})&\:=0,
        \end{array}
    \end{cases}
\end{equation}
%
%
%
where $\boldsymbol{I}$ is the identity matrix, and $p$ is the equilibrium pressure obtained for the mixture thermodynamic closure $p:=p_1=p_2$ given by the second line of \eqref{eq:unevaluated_two_scale_system}. This algebraic equation gives $\alpha_1$ and $p$ respectively as the solution and the value of the equilibrium for given values of $m_1$, $m_2$ and $\alpha_1^d$.
System~\eqref{eq:two_scale-incomp-cons} also admits a supplementary conservation equation
\begin{equation}
    \label{eq:math_entropy_cons}
    \partial_t \tEntropy + \bnabla \bcdot((\tEntropy+p)\vel)
    =0, 
\end{equation}
where the total energy $\tEntropy = \rho \Vert\vel\Vert^2 - \lag$ is a mathematical entropy \citep{godlewski_hyperbolic_1991} for~\eqref{eq:two_scale-incomp-cons}.\footnote{The mathematical entropy of the system is studied as the isothermal limit of the Euler-Fourier model in \cite{serre_structure_2010}.
It is showed to be convex and linked to the physical entropy of the mixture $s$ with $\mathcal{H}=\rho(\varepsilon-Ts) + \tfrac{1}{2}\rho\Vert\vel\Vert^2$ where $\varepsilon = e + Ts$, and $T$ are respectively the internal energy and the temperature of the mixture.}

Remark that $\alpha_1$ is not a variable of the set of conservation laws~\eqref{eq:two_scale-incomp-cons}, but its effective dynamics for smooth solution can be explicitly written by applying the material time derivative to the algebraic equation $p_1(m_1/\alpha_1)=p_2(m_2/(1-\alpha_1-\alpha_1^d))$, and reads
\begin{equation}
    D_t \alpha_1 = \Lambda \bnabla\bcdot\vel,
    \quad
    \text{with}
    \quad
    \Lambda = \alpha_1(1-\alpha_1)\tfrac{\rho_2 c_2^2-\rho_1c_1^2}{\alpha_1\rho_2c_2^2+(1-\alpha_1)\rho_1c_1^2}+\alpha_1^d\tfrac{\alpha_1\rho_2c_2^2}{\alpha_1\rho_2c_2^2+(1-\alpha_1)\rho_1c_1^2}.
\end{equation}
This extends the three-equation model proposed by \cite{chanteperdrix_compressible_2002}, which is recovered when $\VFracd\rightarrow 0$.
\subsection{A dissipative two-pressure model}
\label{sec:two_scale_dissip}
Now that we have set the conservative structure of the model with Hamilton's SAP, we add dissipative processes to the model.
For demonstrative purpose, we propose here to model solely a pressure disequilibrium between the two large-scale phases as it is characteristic of two-phase flow models as in the \cite{baer_two-phase_1986} model. Such a disequilibrium model can encompass several physical phenomena with different timescales (e.g. bubbly flows \cite{drui_small-scale_2019}, stochastic thermodynamic relaxation \cite{perrier_derivation_2021}).
Let us mention that the limit of instantaneous relaxation can also be used to build a numerical scheme that solves the pressure equilibrium of \eqref{eq:two_scale-incomp-cons} \citep{chanteperdrix_compressible_2002}.

In order to make this pressure relaxation a large-scale process that acts on each large-scale phase symmetrically, we define the large-scale volume fractions
\begin{equation}
    \overline{\alpha}_k:=\frac{\alpha_k}{1-\alpha_1^d}, \quad \text{for}\quad k=1,2,
\end{equation}
such that for $\alpha_1^d< 1$,
\begin{equation}
    \alpha_1+\alpha_2+\alpha_1^d=1
    \quad \iff \quad
    \overline{\alpha}_1 + \overline{\alpha}_2=1.
\end{equation}
Then, we relax the pressure equilibrium of \eqref{eq:two_scale-incomp-cons} into
\begin{equation}        
    \label{eq:relax_alpha_bar}
    D_t \overline{\alpha}_1 = \frac{p_1-p_2}{\epsilon},
\end{equation}
where $\epsilon>0$ has the dimension of a dynamic viscosity.
Since the pressure is not unique any more with this pressure relaxation, the momentum equation can be expressed as follows
\begin{equation}
    \partial_t  (\rho \vel) + \bnabla \bcdot  (\rho \vel \otimes  \vel+\overline{p}  \boldsymbol{I})=0,
\end{equation}
where $\overline{p}$ is chosen to provide a signed dissipation of $\tEntropy$.
Following appendix~\ref{app:SAP} 
%
%
\begin{equation}
    \label{eq:def_entropy}
    \begin{aligned}
        \varsigma:=& \partial_t \tEntropy +\bnabla\bcdot((\tEntropy+\overline{p}) \vel) \\
        =& (\overline{p}-\alpha_1p_1-(\alpha_2+\alpha_1^d)p_2)\bnabla\bcdot\vel - (p_1-p_2) \: D_t \alpha_1  \\
        =& (\overline{p}-\overline{\alpha}_1p_1-\overline{\alpha}_2p_2)\bnabla\bcdot\vel - \epsilon (1-\alpha_1^d) (D_t\overline{\alpha}_1)^2.
    \end{aligned}
\end{equation}
Therefore, choosing $\overline{p}:=\overline{\alpha}_1p_1 + \overline{\alpha}_2p_2$ gives a signed mathematical entropy production $\varsigma\le 0$ and the relaxed model reads
\begin{equation}
    \label{eq:two_scale-incomp-diss-2}
    \begin{cases}
        \setlength{\arraycolsep}{0pt}
        \begin{array}{lll}
            \partial_t  m_k &\:+ \:\bnabla \bcdot  (m_k \vel)&\:=0,
            \quad \text{for} \quad
            k=1,2,1^d, \\
            \partial_t  \alpha_1^d &\:+\: \bnabla \bcdot  (\alpha_1^d \vel)&\:=0,\\
            \partial_t  (\rho \vel) &\:+\: \bnabla \bcdot  \left(\rho \vel \otimes  \vel+ \overline{p}\boldsymbol{I}\right)&\:=0, \\
        \end{array}  \\
        D_t \overline{\alpha}_1= \epsilon^{-1}(p_1-p_2).
    \end{cases}
\end{equation}
%
The model (\ref{eq:two_scale-incomp-diss-2}) is obtained for the specific dissipative process~\eqref{eq:relax_alpha_bar} that gives a relaxed model for \eqref{eq:two_scale-incomp-cons}.
Remark that other dissipative processes could have been considered, but are not modelled here, such as drag for two-velocity models \citep{saurel_modelling_2017,gavrilyuk_uncertainty_2020} or turbulent dissipation \cite{saurel_multiphase_2003}.
%
%
%
%
%
\subsection{Discussion of the two-scale models}
\label{sec:two_scale_discussion}
The systems (\ref{eq:two_scale-incomp-cons}) and (\ref{eq:two_scale-incomp-diss-2}) fulfil our first goal of proposing a model that describes simultaneously separated and disperse regimes, and consequently allows a transition between these two regimes.
This mixture couples the different phases with a dissipative pressure relaxation at the large scale while the large-scale phases are coupled with the small-scale one through the constraint on its incompressible volume occupancy.

Let us discuss now the consequences of such a coupling on the mathematical and physical properties of these systems, with a particular interest in their separated regime limit, when $\alpha_1^d\rightarrow 0$, and disperse regime limit, when $\alpha_1\rightarrow 0$.

\input{two_scale_models_eigen}

First, we study two main mathematical properties for our systems: the signed mathematical entropy production $\varsigma$ and the hyperbolicity.
Both of these properties are usually considered as requirements for the system not to be ill-posed \citep{mock_systems_1980,godlewski_hyperbolic_1991,chanillo_remarks_2005}.
A mathematical entropy is identified along with its signed evolution in~\eqref{eq:def_entropy}.
Hyperbolicity of a generic set of $N$ conservation laws on a state vector $\boldsymbol{q}$ is assessed by writing it in its quasi-linear form
\begin{equation}      
    \partial_t \boldsymbol{q} + \boldsymbol{A}(\boldsymbol{q})\partial_x \boldsymbol{q} = 0,
\end{equation}
where $\boldsymbol{A}(\boldsymbol{q})$ is the flux Jacobian.
The hyperbolicity of the conservation laws consists in requiring that the flux Jacobian $\boldsymbol{A}(\boldsymbol{q})$ possesses only real eigenvalues with associated eigenvectors spanning $\mathbb{R}^N$.
%
Denoting $\Delta p = p_1-p_2$, this information is gathered in table~\ref{tab:math_analysis_two_scale} and both models~\eqref{eq:two_scale-incomp-cons} and~\eqref{eq:two_scale-incomp-diss-2} are hyperbolic.
They possess linearly degenerate fields associated to the material wave of velocity $\boldsymbol{u}$ and genuinely non-linear fields associated to sound propagation. From table~\ref{tab:math_analysis_two_scale}, one observes that the presence of the small-scale represented by the incompressible volume fraction $\alpha_1^d > 0$ adds a linearly degenerate field associated with the material wave.

Second, we focus on the physical impact of the incompressible small-scale volume occupation on the global dynamics.
The eigenvalues of the genuinely non-linear fields detailed in table~\ref{tab:math_analysis_two_scale} show that the small scale increases the sound speed by a factor $(1-\alpha_1^d)^{-1}$.
For a dilute disperse regime, \cite{temkin_suspension_2005} showed a similar behaviour when $\alpha_1^d\ll\rho_2/\rho_1^d$. For larger volume fraction, the results differ because of the incompressibility assumption we took in (\ref{hyp:incomp}).
Third, we discuss the separated regime limit $\alpha_1^d \rightarrow 0$ and the disperse regime limit $\alpha_1~\rightarrow~0$.
The models \eqref{eq:two_scale-incomp-cons} and \eqref{eq:two_scale-incomp-diss-2} in the separated limit boil down to those presented in \cite{chanteperdrix_compressible_2002, caro_dinmod_2005}.
In the disperse limit, both models reduce into
\begin{equation}
    \label{eq:disperse_limit_models}
    \begin{cases}
        \setlength{\arraycolsep}{0pt}
        \begin{array}{lll}
            \partial_t  m_k &\:+ \:\bnabla \bcdot  (m_k \vel)&\:=0,
            \qquad
            k=2,1^d, \\
            \partial_t  \alpha_1^d &\:+\: \bnabla \bcdot  (\alpha_1^d \vel)&\:=0,                                                             \\
            \partial_t  (\rho \vel) &\:+\: \bnabla \bcdot  (\rho \vel \otimes  \vel+p_2 \boldsymbol{I})&\:=0.
        \end{array}                                    
    \end{cases}
\end{equation}
The present minimal description of the small-scale is yet insufficient to preserve hyperbolicity when the liquid large-scale disappears.
Very similar phenomenological models for disperse flows with an additional configuration pressure term have been proposed in \cite{raviart_non-conservative_1995, mcgrath_compressible_2016} to recover hyperbolicity.
It requires further investigations to provide a systematic approach generalizing the modelling of such a configuration pressure within Hamilton's SAP.

%% file: two_scale_models_eigen.tex
\begin{table}
    \centering
    \begin{subtable}[t]{.4\textwidth}
        {\tiny
        \setlength{\arraycolsep}{1pt}
        $
            \newlength\cwf
            \newlength\cw
            \newlength\cwt
            \newlength\cwl
            \settowidth{\cwf}{$(m_1,$}
            \settowidth{\cw}{$-\rho_2$}
            \settowidth{\cwt}{$\rho(u-c_W))$}
            \def\arraystretch{1.5}
            \begin{array}{cc wc{160pt}}
                \lambda_k & : & \boldsymbol{r}_k^T\\[-4pt]
                \hline
                u & : & 
                \setlength{\arraycolsep}{10pt}
                \begin{array}{wr{\cwf}wr{\cw}wr{\cwt}}
                    (\rho_1,        & -\rho_2,       & u(\rho_1-\rho_2))      \\[3pt]
                \end{array}\\[3pt]
                & & \\[3pt]
                u+c_W & : &
                \setlength{\arraycolsep}{10pt}
                \begin{array}{wr{\cwf}wr{\cw}wr{\cwt}}
                    (m_1, & m_2, & \rho(u-c_W))\\[3pt]
                \end{array}\\[3pt]
                u-c_W & : &
                \setlength{\arraycolsep}{10pt}
                \begin{array}{wr{\cwf}wr{\cw}wr{\cwt}}
                    (m_1, & m_2, & \rho(u+c_W))\\[3pt]
                \end{array}
            \end{array}
            $}
        \captionsetup{width=\textwidth}
        \caption{3-eq. model for $(m_1, m_2, \rho u)$, i.e. \eqref{eq:two_scale-incomp-cons} in the separated flow limit $\alpha_1^d\rightarrow 0$}
    \end{subtable}
    \hfill
    \begin{subtable}[t]{.5\textwidth}
        {\tiny
            \setlength{\arraycolsep}{1pt}
            $
            \settowidth{\cwf}{$(-(1-\alpha_1^d),$}
            \settowidth{\cw}{$-\rho_2$}
            \settowidth{\cwt}{$\rho(u-c_W^d))$}
            \def\arraystretch{1.5}
            \begin{array}{cc wc{160pt}}
                \lambda_k & : & \boldsymbol{r}_k^T\\[-4pt]
                \hline
                u & : &\left\{ 
                    \setlength{\arraycolsep}{4pt}
                \begin{array}{wr{\cwf}wr{\cw}wr{\cw}wr{\cwt}}
                    (0, &\rho_1,        & -\rho_2,       & u(\rho_1-\rho_2))      \\[3pt]
                    (-(1-\alpha_1^d), &  m_1,    & m_2,       & u(\rho-\rho_1^d))      \\[3pt]
                \end{array}\right.\\[3pt]
                u+c_W^d & : & \hspace{2pt}
                \setlength{\arraycolsep}{4pt}
                \begin{array}{wr{\cwf}wr{\cw}wr{\cw}wr{\cwt}}
                    (\alpha_1^d, & m_1, & m_2, & \rho(u-c_W^d))\\[3pt]
                \end{array}\\[3pt]
                u-c_W^d & : &\hspace{2pt}
                \setlength{\arraycolsep}{4pt}
                \begin{array}{wr{\cwf}wr{\cw}wr{\cw}wr{\cwt}}
                    (\alpha_1^d, & m_1, & m_2, & \rho(u+c_W^d))\\[3pt]
                \end{array}
            \end{array}
            $}
        \captionsetup{width=\textwidth}
        \caption{System (\ref{eq:two_scale-incomp-cons}) for $(\alpha_1^d, m_1, m_2, \rho u)$}
    \end{subtable}\\[10pt]
    \begin{subtable}[t]{.4\textwidth}
        {\tiny
        \setlength{\arraycolsep}{1pt}
        $
            \settowidth{\cwf}{$(0,$}
            \settowidth{\cw}{$-\alpha_2 c_2^2,$}
            \settowidth{\cwt}{$-c_F)$}
            \def\arraystretch{1.5}
            \begin{array}{cc wc{160pt}}
                \lambda_k & : & \boldsymbol{r}_k^T\\[-4pt]
                \hline
                u & : & \left\{
                    \setlength{\arraycolsep}{10pt}
                \begin{array}{wr{\cwf}*{2}{wr{\cw}}wr{\cwt}}
                    (1, & -\frac{\Delta p}{c_1^2},     & -\frac{\Delta p}{c_2^2}, & 0)      \\[3pt]
                    (0, & -\alpha_2 c_2^2, & \alpha_1 c_1^2,  & 0)     \\[3pt]
                \end{array}\right.\\[3pt]
                & & \\[3pt]
                u+c_F & : &\hspace{2pt}
                \setlength{\arraycolsep}{10pt}
                \begin{array}{wr{\cwf}*{2}{wr{\cw}}wr{\cwt}}
                    (0, & \rho_1,              & \rho_2,          & c_F)\\[3pt]
                \end{array}\\[3pt]
                u-c_F & : &\hspace{2pt}
                \setlength{\arraycolsep}{10pt}
                \begin{array}{wr{\cwf}*{2}{wr{\cw}}wr{\cwt}}
                    (0, & \rho_1,              & \rho_2,          & -c_F)
                \end{array}
            \end{array}
            $
            }
        \captionsetup{width=\textwidth}
        \caption{4-eq. model for $(\alpha_1, \rho_1, \rho_2, u)$, i.e. \eqref{eq:two_scale-incomp-diss-2} in the separated flow limit $\alpha_1^d\rightarrow 0$}
    \end{subtable}
    \hfill
    \begin{subtable}[t]{.5\textwidth}
        {\tiny
            \setlength{\arraycolsep}{1pt}
            $
            \settowidth{\cwf}{$(-\Delta p$}
            \settowidth{\cw}{$\rho(c_F^d)^2,$}
            \settowidth{\cwt}{$-(1-\VFracd)\frac{\Delta p}{c_2^2},$}
            \settowidth{\cwl}{$-c_F^d)$}
            \def\arraystretch{1.5}
            \begin{array}{cc wc{160pt}}
                \lambda_k & : & \boldsymbol{r}_k^T\\[-4pt]
                \hline
                u & : & \left\{
                \begin{array}{wr{\cwf}wr{\cw}*{2}{wr{\cwt}}wr{\cwl}}
                    (0, & 1,         & -(1-\VFracd)\frac{\Delta p}{c_1^2},     & -(1-\VFracd)\frac{\Delta p}{c_2^2}, & 0)      \\[3pt]
                    (0, & 0,         & -\alpha_2 c_2^2, & \alpha_1 c_1^2,  & 0)     \\[3pt]
                    (-\Delta p, & \rho(c_F^d)^2,             & 0,         & 0, & 0)\\[3pt]
                \end{array}\right.\\[3pt]
                u+c_F^d & : &\hspace{2pt}
                \begin{array}{wr{\cwf}wr{\cw}*{2}{wr{\cwt}}wr{\cwl}}
                    (\VFracd, & 0,   & (1-\VFracd)\rho_1,              & (1-\VFracd)\rho_2,          & c_F^d)\\[3pt]
                \end{array}\\[3pt]
                u-c_F^d & : &\hspace{2pt}
                \begin{array}{wr{\cwf}wr{\cw}*{2}{wr{\cwt}}wr{\cwl}}
                    (\VFracd, & 0,   & (1-\VFracd)\rho_1,              & (1-\VFracd)\rho_2,          & -c_F^d)
                \end{array}
            \end{array}
            $
        }
        \captionsetup{width=\textwidth}
        \caption{System (\ref{eq:two_scale-incomp-diss-2}) for $(\alpha_1^d, \overline{\alpha}_1, \rho_1, \rho_2, u)$}       
    \end{subtable}\\[10pt]
    {\small
    \hfill
    where
    \hfill
    $
    (c_W^2)^{-1}=\rho\left(\frac{\alpha_1}{\rho_1c_1^2}+\frac{\alpha_2}{\rho_2c_2^2}\right),
    $
    \hfill
    $
    c_F^2 = Y_1c_1^2+Y_2c_2^2,
    $
    \hfill
    $
    (c_W^d)^2 = \frac{c_W^2}{(1-\VFracd)^2},
    $
    \hfill
    $
    (c_F^d)^2 = \frac{c_F^2}{(1-\VFracd)^2}.
    $
    \hfill
    }
    \captionsetup{width=\textwidth}
    \caption{Eigenvalues $\lambda_k$ and eigenvectors $\boldsymbol{r}_k$ for separated and two-scale models.}
    \label{tab:math_analysis_two_scale}
\end{table}

%% file: 2_polydispersion.tex
\section{Small-scale modelling consistent with polydisperse sprays of spherical droplets}
\label{sec:polydisperse}

Now we follow two parallel goals to enrich the small-scale models of the previous section: 1- describing polydisperse sprays in the disperse regime, where the droplets are spherical, in their static minimal surface configuration, and 2- introducing geometric quantities which describe the small-scale for any kind of interfaces in all flow regimes, thus proposing a limit model of a general interface dynamics at small scale.
This modelling strategy has first been introduced for droplets in spherical shape in \cite{essadki_adaptive_2016} and will be referred to in the present contribution as GeoMOM.
This strategy relies on a kinetic modelling of the small scale as a population of droplets.
The resulting model is consequently coherent in the disperse regime where the kinetic approach is valid.
In the mixed regime, the physics validity of the kinetic-based model is limited, but it still remains interpretable since the chosen geometric variables are showed to be defined with both a large-scale modelling of the interface and the small-scale kinetic model.
\subsection{Kinetic-based model in the disperse regime}\label{sec:kin-model}
We start the modelling of the spray by assuming that:
\begin{align}
    & \text{\textbullet\: the small-scale phase is composed of spherical droplets;}  \label{hyp:spheres} \tag{H3a}\\
    & \text{\textbullet\: the droplets do not break nor coalesce.} \label{hyp:no_breakage} \tag{H3b}
\end{align}
Indeed, \eqref{hyp:low_vel_diff} shows that only small velocity differences are locally allowed at the small scale, and droplets are then very close to a spherical shape.
The droplets are then described by a static minimal surface, which remains spherical. 
The kinetic modelling of the disperse small-scale relies on the Number Density Function (NDF) $f$ that counts the number of droplets within a small volume of the phase space around a point of the space-time domain.
For generality purpose, we adopt a description for compressible inclusions whose sizes can change in time under either mass exchange or compressibility processes.
Therefore, we consider the NDF $f(\boldsymbol{x}, t, \boldsymbol{v}, \widehat{m})$ to be a distribution of velocities $\boldsymbol{v}$ and droplet masses $\widehat{m}$ instead of a size parameter such as radius or surface.
We also denote $f_{\boldsymbol{v},m}$ for compactness to underline the phase-space dependencies only.
Then, the equation of evolution of the NDF expresses the conservation of the droplets in both the real and phase spaces \citep{williams_spray_1958,marchisio_computational_2013} and reads 
\begin{equation}
    \label{eq:GPBE_m}
    \partial_t f_{\boldsymbol{v},m}
    + \bnabla_{\boldsymbol{x}}\bcdot(f_{\boldsymbol{v},m} \boldsymbol{v})
    + \bnabla_{\boldsymbol{v}}\bcdot(\boldsymbol{F}(\boldsymbol{v}, \widehat{m})f_{\boldsymbol{v},m})
    + \partial_{\widehat{m}}(R_{m}(\boldsymbol{v}, \widehat{m})f_{\boldsymbol{v},m})=\Gamma,
\end{equation}
where $\boldsymbol{F}$, and $R_{m}$ are unclosed rates of change of the velocity and the mass corresponding to body forces and mass transfer, while $\Gamma=\Gamma_{bu}+\Gamma_{coal}$ accounts for source terms such as break-up and coalescence phenomena. (\ref{hyp:no_breakage}) and (\ref{hyp:ls_no_mass}) respectively imply $\Gamma=0$, and $R_{m}=0$.
Moreover, the distribution of velocities is out of the scope of this work as assumed by (\ref{hyp:ls_shared_vel}) (see \cite{vie_size-velocity_2013} for an Eulerian model coupling the effects of size and velocity distributions).
Therefore, we discard the velocity dependency by considering $n_m = \int_{\boldsymbol{v}} f_{\boldsymbol{v}, m} d\boldsymbol{v}$, the mass-based NDF in the limit of a vanishing Stokes number $St \rightarrow 0$, i.e. when the inertial timescale of inclusions is negligible compared to the timescale of the flow. 
Following the works of \cite{jabin_various_2002, massot_eulerian_2007}, the dynamics of $n_m$ reads
\begin{equation}
    \label{eq:PBE_m}
        \partial_t n_{m} + \bnabla_{\boldsymbol{x}}\bcdot(n_{m} \vel)=0,
\end{equation}
where $\vel$ is the unique velocity of every phase as defined by~\eqref{eq:unique_vel}.
\subsection{Geometric Method Of Moments : definition and application to the incompressible spray}
\label{sec:geomom}
GeoMOM aims at enriching the models \eqref{eq:two_scale-incomp-cons}-\eqref{eq:two_scale-incomp-diss-2} with quantities describing the geometry of the interface at the small-scale.
This method introduced in \cite{essadki_adaptive_2016, essadki_high_2018, essadki_statistical_2019} is a reduced-order moment model for the polydisperse droplet distribution where the chosen moments are geometric quantities that can be defined for any flow regime and interface topology.


First, let us define those quantities.
We consider a surface $\mathcal{S}$ defined by mapping a set $\mathcal{U}\subset \mathbb{R}^2$ onto $\mathcal{S}\subset\mathbb{R}^3$ such that we denote $A(u,v)\:dudv$ the infinitesimal surface element over $\mathcal{S}$. 
Then, the surface area and a surface-average operator are defined on $\mathcal{S}$ by
\begin{equation}\label{eq:average_op}
    S:= \int_{\mathcal{U}} A(u,v)\:dudv,
    \qquad
    \left<\:\cdot\:\right>:=\frac{1}{S}\int_{\mathcal{U}}(\cdot)\: A(u,v)\:dudv.
\end{equation}
Considering the surface of the droplets in the two-scale mixture as described in Section~\ref{sec:two-scale}, the sum of their areas defines the small-scale interface area density $\Sigma$.
Thanks to~\eqref{eq:average_op}, one can also define surface-average Gauss and mean curvatures $\SigG$ and $\SigH$ as in~\cite{drew_mathematical_1983,pope_evolution_1988}.
We use here these quantities to enhance the description of the disperse small scale, but their definitions are \textit{a priori} independent of the surface geometry (see an application of these geometric quantities to the interface resulting from for the collision of two spherical droplets in appendix \ref{app:geomom_recon}).
This property is particularly convenient as it indicates that the kinetic-based model can be interpreted as a limit case for the mixed regime, and we discuss in conclusion how it allows coupling of the interface geometry between scales.

Second, we reduce the complexity of the spray dynamics, modelled by the distribution $n_m$ given in~\eqref{eq:PBE_m}, by introducing a finite set of moments indexed by $\mathcal{I}$
\begin{equation}
    M_i^m := \int_{\widehat{m}} \widehat{m}^i \: n_m \: d\widehat{m},
    \qquad
    i\in \mathcal{I} \text{ finite }\subset\mathbb{N}.
\end{equation}
These scalars retain statistical information about the NDF and their dynamics can be obtained by integrating \eqref{eq:PBE_m} against the corresponding monomials,
\begin{equation}
    \label{eq:cons_mom_m}
    \partial_t M_i^m + \bnabla\bcdot(M_i^m\vel)=0,
    \qquad
    i\in\mathcal{I}.
\end{equation}
The method of moments then gives a reduced-order model for the small-scale spray in comparison with \eqref{eq:PBE_m}.
In general, the equations are unclosed, and a reconstruction of the NDF based on the selected moments must be provided.
The selection of these moments is not obvious and is usually motivated on the basis of mathematical properties of the resulting moment model.

Finally, the specificity of GeoMOM consists in the selection of moments related to geometric quantities defined with~\eqref{eq:average_op} to construct a model interpretable even out of the disperse regime.
For the spherical droplets described by $n_m$, the local Gauss and mean curvatures on the sphere are constant and equal respectively to $R^{-1}$ and $R^{-2}$ where $R=(3m/(4\pi\rho_1^d))^{1/3}$ is the radius of the sphere of density $\rho_1^d$ and mass $m$.
Therefore, considering a population of spherical droplets, we express the geometric quantities $\Sigma$, $\Sigma\left<G\right>$ and $\Sigma\left<H\right>$ as moments of the distribution $n_m$
\begin{equation}
    \label{eq:relations_geom_moments}
    \begin{aligned}
        \SigG &= \int_{\widehat{m}} 4\pi && && n_m \:d\widehat{m}&&=4\pi && M^m_0,\\
        \SigH &= \int_{\widehat{m}} 4\pi \left(\frac{3}{4\pi \rho_1}\right)^{1/3} && \widehat{m}^{1/3} && n_m \:d\widehat{m}&&=4\pi \left(\frac{3}{4\pi \rho_1^d}\right)^{1/3} && M^m_{1/3},\\
        \Sigma&= \int_{\widehat{m}} 4\pi \left(\frac{3}{4\pi \rho_1^d}\right)^{2/3} && \widehat{m}^{2/3} && n_m \:d\widehat{m}&&=4\pi \left(\frac{3}{4\pi \rho_1^d}\right)^{2/3} && M^m_{2/3}.
    \end{aligned}
\end{equation}
Remark the special role of $\SigG$ proportional to the $0^{th}$-order moment of $n_m$.
This results from the Gauss-Bonnet theorem \cite{kreyszig_differential_1991}, which indicates a geometric invariant $S\widetilde{G}=4\pi$ for continuous deformations of the sphere, where $\widetilde{(\cdot)}$ denotes the surface average as~\eqref{eq:average_op} applied to a unique droplet.
It is also showed in \cite{essadki_statistical_2019} that such an invariant enables to write the dynamics of geometric quantities $\Sigma$, $\Sigma\left<\:\cdot\:\right>$ independently of the flow regime.
Furthermore, the small-scale volume fraction is not surface-related in general, but in the specific case of sperical droplets, it is also linked to a moment of $n_m$ with
\begin{equation}
    \label{eq:alpha_1d_mom_rel}
    \alpha_1^d=\int_{\widehat{m}} \frac{1}{\rho_1^d} \widehat{m} \: n_{m} \:d\widehat{m} = \frac{1}{\rho_1^d} M^m_1.
\end{equation}
From \eqref{eq:relations_geom_moments} and \eqref{eq:alpha_1d_mom_rel}, we select $\mathcal{I}=\{0,1/3,2/3,1\}$ in \eqref{eq:cons_mom_m} to get the dynamics of the geometric quantities
\begin{equation}
    \label{eq:geom_dynamics_compressible}
    \begin{cases}
        \setlength{\arraycolsep}{0pt}
        \begin{array}{lll}
            \partial_t (\Sigma\left<G\right>) &\:+ \: \bnabla \bcdot (\Sigma\left<G\right> \vel)&\:=0,\\
            \partial_t ((\rho_1^d)^{1/3}\Sigma\left<H\right>) &\:+ \: \bnabla \bcdot ((\rho_1^d)^{1/3}\Sigma\left<H\right> \vel)&\:=0,\\
            \partial_t ((\rho_1^d)^{2/3}\Sigma) &\:+ \: \bnabla \bcdot ((\rho_1^d)^{2/3}\Sigma \vel)&\:=0,\\
            \partial_t m_1^d &\:+ \: \bnabla \bcdot (m_1^d \vel)&\:=0.
        \end{array}
    \end{cases}
\end{equation}
Up to here, the model has been derived for compressible spherical inclusions.
We focus back on the description of a spray of droplets by assuming the incompressibility of the small scale thanks to the constraint $D_t \rho_1^d = 0$.
Then, the geometric variables are governed by
\begin{equation}
    \label{eq:conserved_geom_var}
    \begin{cases}
        \setlength{\arraycolsep}{0pt}
        \begin{array}{lll}
            \partial_t (\Sigma\left<G\right>) &\:+ \: \bnabla \bcdot (\Sigma\left<G\right> \vel) &\:= 0,\\
            \partial_t (\Sigma\left<H\right>) &\:+ \: \bnabla \bcdot (\Sigma\left<H\right> \vel) &\:= 0,\\
            \partial_t \Sigma &\:+ \: \bnabla \bcdot (\Sigma \vel) &\:= 0,\\
            \partial_t \alpha_1^d &\:+ \: \bnabla \bcdot (\alpha_1^d \vel) &\:= 0.
        \end{array}
    \end{cases}
\end{equation}
System~\eqref{eq:conserved_geom_var} corresponds to the system with surface-based moments of \cite{essadki_high_2018} when no evaporation nor condensation is accounted for.
Indeed, when the droplets are incompressible, the mass-based NDF $n_m$ relates to the surface-based NDF $n_{S}$ defined by
\begin{equation}
    \label{eq:surface_based_ndf_def}
    n_S(\boldsymbol{x}, t, \widehat{S}):=n_m(\boldsymbol{x}, t, \widehat{m})\delta(\widehat{m}-m(\widehat{S})),
    \qquad
    m(\widehat{S})=\rho_1^d\frac{\widehat{S}}{3\sqrt{4\pi}}.
\end{equation}
As $\rho_1^d$ is a constant along the streamlines, there is no variation of surface area for the droplets and the dynamics is then driven by
\begin{equation}
    \label{eq:PBE_S}
        \partial_t n_S + \bnabla_{\boldsymbol{x}}(n_S \vel)=0.
\end{equation}
Again, when the droplets have the same density, the geometric quantities are also expressed through the moments $M_i^S$ of $n_S$,
\begin{equation}
    \label{eq:geom_var_moments_rel_S}
    \Sigma\left<G\right> = 4\pi M^S_0,
    \quad
    \Sigma\left<H\right> = \sqrt{4\pi} M^S_{1/2},
    \quad
    \Sigma = M^S_1,
    \quad
    \alpha_1^d = \frac{1}{3 \sqrt{4\pi}} M^S_{3/2}.
\end{equation}
The change of variables~\eqref{eq:surface_based_ndf_def} only modifies the dimensions and orders of the moments.
Consequently, the four initial half-integer moments can be recovered as in \cite{essadki_high_2018}.
In the following, we will stick with the surface-based moments for better comparison with the work of Essadki.
In the end, the relations \eqref{eq:geom_var_moments_rel_S} between moments and geometric quantities provide a reduced-order model of the small-scale disperse phase (see appendix \ref{app:geomom_recon} for an illustration of the reduced-order modelling applied to droplets resulting from the collision of two spherical droplets). 
\subsection{A two-scale model with a polydisperse small-scale model}
The two-scale models derived in Section \ref{sec:two-scale} thanks to Hamilton's SAP can now be enhanced by incorporating the new geometric quantities in the Lagrangian and specifying whether these parameters are constrained by new conservation laws.
We add then a small-scale capillarity energy $\sigma\Sigma$ along with the conservation constraints given by \eqref{eq:conserved_geom_var}.
Let us introduce a new variable $z=\Sigma/m_1^d$, which is transported, that is $D_t z = 0$. As the large-scale capillarity is neglected (\ref{hyp:ls_no_capillarity}), the energies associated with the large scales of the flow are left unchanged whereas the small-scale Lagrangian $\mathcal{L}_1^d$ becomes
\begin{equation}
    \label{eq:lag_small_scale_capillarity}
    \lag_1^d\left(m_1^d, \rho_1^d, z:=\Sigma/m_1^d, \vel\right):=\frac{1}{2}m_1^d \vert\vel\vert^2
    -m_1^d e_1(\rho_1^d)
    -\sigma m_1^d z.
\end{equation}
Eventually, the system resulting from Hamilton's SAP (see details in appendix \ref{app:SAP_polydisperse}) together with the constraints~\eqref{eq:conserved_geom_var} yields
\begin{equation}
    \label{eq:two_scale_model_polydisperse}
    \begin{cases}
        \setlength{\arraycolsep}{0pt}
        \begin{array}{lll}
            \partial_t  m_k &\:+ \:\bnabla \bcdot  (m_k \vel)&\: =0,
            \qquad
            k=1,2,1^d, \\
            \partial_t  \alpha_1^d &\:+\: \bnabla \bcdot  (\alpha_1^d \vel)& \: =0,                                                             \\
            \partial_t \Sigma &\:+\:\bnabla \bcdot(\Sigma\vel)&\: =0,\\
            \partial_t  (\Sigma\left<G\right>) &\:+\: \bnabla \bcdot  (\Sigma\left<G\right> \vel) &\: =0,                          \\
            \partial_t (\Sigma\left<H\right>) &\:+\: \bnabla \bcdot(\Sigma\left<H\right>\vel)&\: =0,\\
            \partial_t (\rho \vel)&\:+\: \bnabla \bcdot (\rho \vel \otimes \vel+p\boldsymbol{I})&\:=0,
        \end{array}
    \end{cases}
\end{equation}
with $p:=p_1=p_2$. Let us remark that the new small-scale capillarity term in the Lagrangian does not alter momentum flux in this formulation compared to \eqref{eq:two_scale-incomp-cons}.
Similarly to the previous models but with the updated definition of the Lagrangian $\mathcal{L}$, this model also admits a supplementary conservation equation
\begin{align}
    \label{eq:math_entropy_cons_poly}
    \partial_t \tEntropy + \bnabla \bcdot((\tEntropy+p)\vel)
    &=0,
    &
    \tEntropy &= \sum_k m_k \Vert\vel\Vert^2 - \lag,
\end{align}
and a pressure relaxation can be built as in \eqref{eq:two_scale-incomp-diss-2}.
Both~\eqref{eq:two_scale_model_polydisperse} and its relaxed version are hyperbolic with similar structure of eigenvalues and eigenvectors as in table~\ref{tab:math_analysis_two_scale} with new fields transported at velocity $\vel$ as the additional variables are associated with transport equation at velocity $\vel$.
If needed, interaction between scales can be accounted for by involving geometric quantities in the pressure relaxation, and such an extension is discussed in the conclusion.
%

%% file: 3_oscillation.tex
\section{Polydisperse and vibrational spray small-scale model}
\label{sec:kinetic-model-oscill}
\begin{figure}
    \centering
    \includegraphics[width=.9\textwidth]{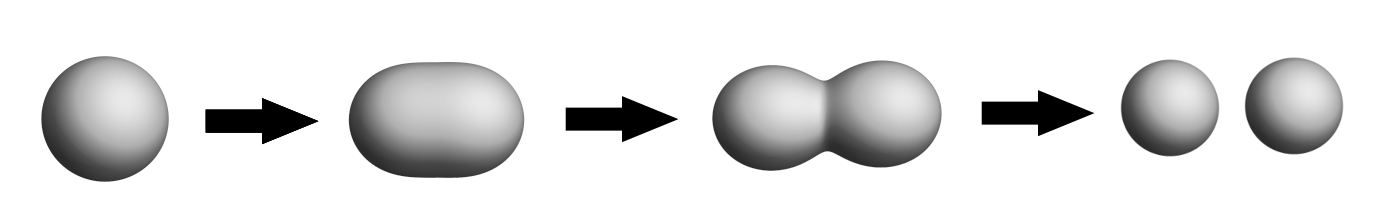}
    \caption{Sphere deformed with the second axisymmetric spherical harmonic and vibrational breakup modelled by a critical length for the smallest radius.}
    \label{fig:vibrational}
\end{figure}

In the previous model, no dynamics have been associated with the geometric variables as they represent a spray of spherical droplets, that is a static minimal surface.
Here, we propose to enhance this model by associating a dynamics to the droplet geometry.
In the context of the small velocity differential given by (\ref{hyp:low_vel_diff}), we add an oscillating motion to the droplets around their spherical shapes.
Such a model is sufficiently simple to identify the associated energies while being a first step towards the modelling of break-up as proposed in \cite{orourke_tab_1987, amsden_kiva-ii_1989} (see figure \ref{fig:vibrational}).
Moreover, this model introduces with GeoMOM non-trivial dynamics for geometric quantities such as interface area density.
\subsection{Kinetic-based model in the disperse regime}
Let us model the spray of oscillating droplets by first discarding assumption \eqref{hyp:spheres} of spherical shapes, and rather assume:
\begin{align}
    & \text{\textbullet\: the droplets' internal flow is irrotational;
    } \label{hyp:osc_irrot} \tag{H4a} \\
    & \text{\textbullet\: the amplitude of the interface deformation is small;
    }  \label{hyp:osc_small_amp} \tag{H4b}\\
    & \text{\textbullet\: the deformed droplets are either randomly orientated or they all } \notag \\
    & \text{orientate to the same privileged direction;} \label{hyp:orientation} \tag{H4c}\\
    & \text{\textbullet\: the droplets' interface is deformed along the second 
    } \notag \\
    & \text{axisymmetric spherical harmonic.} \label{hyp:osc_sec_harm} \tag{H4d}
\end{align}
The dynamics of a droplet satisfying (\ref{hyp:osc_irrot}) and (\ref{hyp:osc_small_amp}) has been studied thoroughly \citep{john_w_strutt_3rd_baron_rayleigh_vi_1879,prosperetti_viscous_1977,plumacher_non-linear_2020}.
The choice of orientation (\ref{hyp:orientation}) of the deformation depends on the physics, whether it is isotropic or not, and it is not discussed here.
Denote $\boldsymbol{r}$ the position of the droplet interface in spherical coordinates $(r,\theta,\phi)$ and $(\boldsymbol{e}_r$, $\boldsymbol{e}_{\theta}$, $\boldsymbol{e}_{\phi})$ the spherical orthonormalized basis.
Following (\ref{hyp:osc_sec_harm}), we denote $Y_2(\theta) = \sqrt{5}/(4\sqrt{\pi})(3\cos^2\theta-1)$ the second axisymmetric spherical harmonic, and $R_0$ the radius of the non-deformed spherical droplet, such that the position writes
\begin{equation}
    \label{eq:osci_position_vec}
    \boldsymbol{r} = (R_0 + x_2 Y_2) \boldsymbol{e}_r,
\end{equation}
where $x_2$ denotes the amplitude of the deformation.
The dynamics of this motion follows the harmonic oscillation
\begin{equation}
    \label{eq:small_scale_oscillator}
    \ddot{x}_2+\omega^2 x_2 = 0,
    \qquad
    \omega^2 = 8\frac{\sigma}{\rho_1^d R_0^3}=\tilde{\omega}^2S_0^{-3/2},
\end{equation}
where $\tilde{\omega}^2=(8(4\pi)^{3/2}\sigma/\rho_1^d)$ is a constant along the streamlines.
The dynamics of the droplets is then characterized by $(S_0,x_2,\dot{x}_2)$.
For computational reasons, we introduce $\chi = (2/S_0)^{1/2}x_2$ which satisfies the same dynamics~\eqref{eq:small_scale_oscillator} as $x_2$. Assuming \eqref{hyp:no_breakage} and \eqref{hyp:ls_no_mass}, the balance equation for the NDF $n_{\xi}$ in the phase-space $\widehat{\boldsymbol{\xi}}=(\widehat{S}_0, \widehat{\chi}, \widehat{\dot{\chi}})$ yields
\begin{equation}
    \label{eq:unclosed_osc_PBE}
    \partial_t n_{\xi}
    + \bnabla_{\boldsymbol{x}} \bcdot (n_{\xi}\vel)
    + \partial_{\widehat{\chi}} (R_{\chi}n_{\xi})
    + \partial_{\widehat{\dot{\chi}}}(R_{\dot{\chi}}n_{\xi})=0,
\end{equation}
where $R_{\chi}$ and $R_{\dot{\chi}}$ are rates of change that remain to be closed.
This model can be seen as the particular case satisfying \eqref{hyp:ls_shared_vel}-\eqref{hyp:ls_no_mass}-\eqref{hyp:no_breakage}-\eqref{hyp:isothermal} of the kinetic model considered by \cite{orourke_tab_1987, amsden_kiva-ii_1989}.
With the oscillator model (\ref{eq:small_scale_oscillator}), we close~\eqref{eq:unclosed_osc_PBE} by fixing $R_{\chi} = \widehat{\dot{\chi}}$ and $R_{\widehat{\dot{\chi}}} = -\tilde{\omega}^2\widehat{S}_0^{-3/2}\widehat{\chi}$,
\begin{equation}
    \label{eq:WBE_poly_vib}
    \partial_t n_{\xi}
    + \bnabla_{\boldsymbol{x}} \bcdot (n_{\xi}\vel)
    + \partial_{\widehat{\chi}} (\widehat{\dot{\chi}} n_{\xi})
    + \partial_{\widehat{\dot{\chi}}}(-\tilde{\omega}^2\widehat{S}_0^{-3/2}\widehat{\chi} n_{\xi})=0.
\end{equation}
\subsection{Modelling a spray of asynchronously oscillating droplets}
\label{sec:disorg_osc}
With the new PBE \eqref{eq:WBE_poly_vib}, GeoMOM leads to a non-trivial closure problem and the geometric quantities chosen in \eqref{eq:relations_geom_moments} and \eqref{eq:alpha_1d_mom_rel} cannot model an arbitrary distribution.
With such a choice, we propose in this section to approximate the distribution with 
\begin{equation}
    \label{eq:unifrom_closure}
    n_{\xi}(\boldsymbol{x}, t, \widehat{S}_0, \widehat{\chi}, \widehat{\dot{\chi}})
    = n_{S_0}(\boldsymbol{x}, t, \widehat{S}_0)
    \frac{1}{\vert\mathcal{E}(\widehat{S}_0)\vert}
    \mathbb{1}_{\mathcal{E}(\widehat{S}_0)}(\widehat{\chi}, \widehat{\dot{\chi}}),
\end{equation}
where the amplitudes and their rates of change $(\widehat{\chi}, \widehat{\dot{\chi}})$ are uniformly distributed on a compact space $\mathcal{E}\subset\mathbb{R}^2$ of area $\vert\mathcal{E}(\widehat{S}_0)\vert$ that allows a maximal energy for a given droplet size $\widehat{S}_0$.
This approximation of the NDF discards situations where droplets oscillate synchronously, and consequently a macroscopic oscillation motion of the spray.
\subsubsection{GeoMOM based on the classical surface-average operator}
We apply GeoMOM with the same surface-average operators $\left<\:\cdot\:\right>$ and $\widetilde{(\cdot)}$ defined in Section \ref{sec:geomom}, and geometric quantities $\SigG$, $\SigH$, $\Sigma$, $\alpha_1^d$.
We establish then the relations between these geometric quantities and moments of the NDF in the context of a disperse regime,
\begin{equation}
  \label{eq:def_oscillation_quantities}
  \SigG = \int_{\widehat{\boldsymbol{\xi}}}
  S\widetilde{G} \:n_{\xi}\: d\widehat{\boldsymbol{\xi}},
  \quad
  \SigH = \int_{\widehat{\boldsymbol{\xi}}}
  S\widetilde{H} \:n_{\xi}\: d\widehat{\boldsymbol{\xi}},
  \quad
  \Sigma = \int_{\widehat{\boldsymbol{\xi}}}
  S \:n_{\xi}\: d\widehat{\boldsymbol{\xi}},
  \quad
  \alpha_1^d = \int_{\widehat{\boldsymbol{\xi}}}
  V \:n_{\xi}\: d\widehat{\boldsymbol{\xi}}.
\end{equation}
When the oscillations are small, the integrands are functions of $\widehat{\boldsymbol{\xi}}$ (see appendix \ref{app:diff_geom}) and write
\begin{equation}
    \label{eq:droplet_geometry}
    S\tilde{G}=4\pi,
    \quad
    S\tilde{H} - S_0\tilde{H}_0 = \sqrt{4\pi} S_0^{1/2}\chi^2,
    \quad
    V = \frac{1}{3\sqrt{4\pi}} S_0^{3/2},
    \quad
    S - S_0 = S_0 \chi^2,
\end{equation}
where $\tilde{H}_0=\sqrt{4\pi}S_0^{-1/2}$ is the surface averaged mean curvature when the droplet is a sphere \textit{i.e.} $\chi=0$.
Remark that $S\tilde{G}$ and $\alpha_1^d$ are constant despite the oscillation thanks to the Gauss-Bonnet theorem \cite{kreyszig_differential_1991} and the incompressibility assumption.
Finally, relations between geometric quantities and moments $M_{i,j,k}^{\xi}:=\int_{\widehat{\boldsymbol{\xi}}} \widehat{S}^i \widehat{\chi}^j  \widehat{\dot{\chi}}^k \:n_{\xi}\:d\widehat{\boldsymbol{\xi}}$ of $n_{\xi}$ are obtained 
\begin{equation}
    \label{eq:geomom_osc_rel}
    \begin{aligned}
        \SigG &= 4\pi M_{0,0,0}^{\xi},
        &\SigH &= \sqrt{4\pi} (M_{1/2, 0, 0}^{\xi} + M_{1/2,2,0}^{\xi}),\\
        \Sigma &= M_{1,0,0}^{\xi}+M_{1,2,0}^{\xi},
        &\alpha_1^d &= \frac{1}{3\sqrt{4\pi}}M_{3/2,0,0}^{\xi}.
    \end{aligned}
\end{equation}
These relations extend the ones of \eqref{eq:geom_var_moments_rel_S} with moments dedicated to the oscillatory dynamics.
Such decomposition leads us to define and choose the following geometric quantities and moments for our model
\begin{equation}
    \label{eq:geomom_osc_rel_decomp}
    \begin{aligned}
        \SigH_0 &:= \sqrt{4\pi} M_{1/2, 0, 0}^{\xi},
        &\Delta\SigH &:= \sqrt{4\pi} M_{1/2,2,0}^{\xi},\\
        \Sigma_0 &:= \sqrt{4\pi} M_{1, 0, 0}^{\xi},
        &\Delta\Sigma &:= \sqrt{4\pi} M_{1,2,0}^{\xi},
    \end{aligned}
\end{equation}
instead of just $\Sigma$ and $\SigH$ which can be reconstructed with \eqref{eq:geomom_osc_rel}.
Integrating \eqref{eq:WBE_poly_vib} against $(1,\widehat{S}_0^{1/2},\widehat{S}_0,\widehat{S}^{3/2}_0,\widehat{S}^{1/2}_0\widehat{\chi}^2,\widehat{S}_0\widehat{\chi}^2)$ provides
\begin{equation}
    \label{eq:dyn_geom_var_S}
    \begin{cases}
        \setlength{\arraycolsep}{0pt}
        \begin{array}{lll}
            \partial_t (\SigG) &\:+ \: \bnabla \bcdot (\SigG \vel) &\:= 0,\\
            \partial_t (\SigH_0) &\:+ \: \bnabla \bcdot (\SigH_0 \vel) &\:= 0,\\
            \partial_t \Sigma_0 &\:+ \: \bnabla \bcdot (\Sigma_0 \vel) &\:= 0,\\
            \partial_t \alpha_1^d &\:+ \: \bnabla \bcdot (\alpha_1^d \vel) &\:= 0,\\
            \partial_t (\Delta\SigH) &\:+ \: \bnabla \bcdot (\Delta\SigH \vel) &\:= 2 \sqrt{4\pi} M_{1/2,1,1}^{\xi},\\
            \partial_t (\Delta\Sigma) &\:+ \: \bnabla \bcdot (\Delta\Sigma \vel) &\:= 2 M_{1,1,1}^{\xi},\\
        \end{array}
    \end{cases}
\end{equation}
We obtained that $\SigG$, $\SigH_0$, $\Sigma_0$ and $\alpha_1^d$ are conserved similarly as \eqref{eq:conserved_geom_var} with two additional equations for the oscillatory components $\Delta\SigH$ and $\Delta\Sigma$.
\subsubsection{Energies of the spray}
In the context of two-scale modelling with Hamilton's SAP, we are specifically interested in defining the energies of the spray with the geometric quantities.
For the oscillatory motion described by \eqref{eq:small_scale_oscillator}, the kinetic and potential energies of a droplet can be expressed as function of $\boldsymbol{\xi}$ \citep[appendix II]{john_w_strutt_3rd_baron_rayleigh_vi_1879}
\begin{align}
    E^{kin,1d} = \frac{1}{2} \frac{\rho_1^d}{4(4\pi)^{3/2}}S_0^{5/2}\dot{\chi}^2,
    &&
    E^{pot,1d} = \sigma S = \sigma S_0 + \sigma S_0 \chi^2.
\end{align}
It is then straightforward to obtain the energies of the spray $E^{kin,d}$ and $E^{pot,d}$ from moments of $n_{\xi}$ by integrating the expressions above,
\begin{align}
    \label{eq:osc_energies}
    E^{kin,d} = \frac{1}{2} \frac{\rho_1^d}{4(4\pi)^{3/2}}M_{5/2,0,2}^{\xi},
    &&
    E^{pot,d} = \sigma \Sigma = \sigma M^{\xi}_{1,0,0} + \sigma M^{\xi}_{1,2,0}.
\end{align}
Remark that the moment $M_{5/2,0,2}^{\xi}$ is not linked to any of the selected geometric quantities of the model.
We can find a closure for this moment using the approximation of $n^{\xi}$ given in \eqref{eq:unifrom_closure}.
It requires to provide a definition for $\mathcal{E}$ which is chosen following the break-up criterion of \cite{orourke_tab_1987}.
Therefore, we authorize the droplets to oscillate with an energy lower than a fraction $c\in[0,1]$ of the maximal energy $E_{max}$ before break-up,
\begin{equation}
    \mathcal{E}(\widehat{S}_0):=\left\{
        (\widehat{\chi},\widehat{\dot{\chi}})\in\mathbb{R}^2
        \text{ s.t. }
        E^{kin,1d}(\widehat{S}_0, \widehat{\chi}, \widehat{\dot{\chi}}) + E^{pot,1d}(\widehat{S}_0, \widehat{\chi}, \widehat{\dot{\chi}})\le c E_{max}(\widehat{S}_0)
        \right\}.
\end{equation}
The break-up energy $E_{max}(S_0)$ corresponds to a deformation up to an equatorial radius reaching half the value of the spherical radius
\begin{equation}
    R_{eq} := \Vert\boldsymbol{r}(\pi/2)\Vert = \frac{R_0}{2}.
\end{equation}
Then, the subset $\mathcal{E}$ boils down to $\mathcal{E}=\{\chi^2 + (\dot{\chi}/\omega)^2 \le 2c/5\}$ whose area is $\vert\mathcal{E}\vert = \frac{2}{5} \pi \omega c$.
We can now close the expression of $E^{kin,d}$, $E^{pot,d}$ in~\eqref{eq:osc_energies} and find the dynamics of the parameter $c$ using~\eqref{eq:dyn_geom_var_S} and the approximation of the NDF \eqref{eq:unifrom_closure}.
It yields
\begin{equation}
    \begin{gathered}
        E^{kin,d} = \frac{1}{10} c \sigma \Sigma_0,
        \quad
        E^{pot,d} = \sigma\Sigma_0 + \frac{1}{10} c \sigma \Sigma_0,\quad
        D_t c = 0.
    \end{gathered}
\end{equation}
Remark that, as $(\chi,\dot{\chi})$ are uniformly distributed in $\mathcal{E}$, the mechanical energy is evenly distributed between kinetic and potential energies.
Regarding the dynamics of $c$, the energy of the oscillation is advected along the streamline.
Moreover, using the relation $\Delta\Sigma = M^{\xi}_{1,2,0}$, one can replace $c$ by geometric quantities with $c=10\Delta\Sigma/\Sigma_0$. 
\subsubsection{Two-scale model with the small-scale spray model of asynchronous droplets}
Energies related to the small-scale oscillation are negatively signed in Hamilton's SAP as the kinetic energy is not a quadratic form of any kind of velocity.
Indeed, the energy of the droplets is here considered as a whole energetic contribution without following their specific dynamics.
We keep then the expression of the Lagrangian defined in~\eqref{eq:lagrangian_two_scale}, where we only modify the small-scale energies of $\mathcal{L}_1^d$ following
\begin{equation}
    \label{eq:lagrangian_small_scale_unif}
    \mathcal{L}_1^d := 
    \frac{1}{2}m_1^d \Vert\vel\Vert^2
    - m_1^d e_1(\rho_1^d)
    - \sigma \Sigma_0 \left(1 + \frac{c}{5}\right).
\end{equation}
The constraints are the same as in Section \ref{sec:polydisperse} with the additional advection constraint $D_t c=0$. The derivation of the two-scale mixture's dynamics with Hamilton's SAP is very similar to the one of appendix \ref{app:SAP_polydisperse} and yields
\begin{equation}
    \label{eq:two_scale_model_osc_unif}
    \begin{cases}
        \setlength{\arraycolsep}{0pt}
        \begin{array}{lll}
            \partial_t  m_k &\:+ \:\bnabla \bcdot  (m_k \vel)&\: =0,
            \qquad
            k=1,2,1^d, \\
            \partial_t  \alpha_1^d &\:+\: \bnabla \bcdot  (\alpha_1^d \vel)& \: =0,                                                             \\
            \partial_t \Sigma_0 &\:+\:\bnabla \bcdot(\Sigma_0\vel)&\: =0,\\
            \partial_t (\Delta\Sigma) &\:+\: \bnabla \bcdot(\Delta\Sigma\vel)&\: =0,\\
            \partial_t  (\SigG) &\:+\: \bnabla \bcdot  (\SigG \vel) &\: =0,                          \\
            \partial_t (\Sigma\left<H\right>_0) &\:+\: \bnabla \bcdot(\Sigma\left<H\right>_0\vel)&\: =0,\\
            \partial_t (\rho \vel)&\:+\: \bnabla \bcdot (\rho \vel \otimes \vel+p\boldsymbol{I})&\:=0,
        \end{array}
    \end{cases}
\end{equation}
with $p:=p_1=p_2$, and the additional equation on total energy as a mathematical entropy for~\eqref{eq:two_scale_model_osc_unif},
\begin{equation}
    \partial_t \tEntropy + \bnabla \bcdot ((\tEntropy+p)\vel)=0,
    \qquad
    \tEntropy = \sum_k m_k \Vert\vel\Vert^2-\lag.
\end{equation}
Now considering dissipation processes, we could also add the pressure relaxation as in the previous two-scale models.
We propose here to focus on the dissipation associated to the small-scale oscillatory dynamics.
Indeed, the oscillation motion eventually decreases and the associated energy dissipates into thermal energy or small-scale kinetic energy of the gas phase.
Each of these last two energies are not modelled here, so the system loses this energy, and we only consider the following source term $R_c$ on the dynamics of $c$,
\begin{equation}
    D_t c = R_c.
\end{equation}
The mathematical entropy production writes
\begin{equation}
    \varsigma:=\partial_t \tEntropy + \bnabla \bcdot ((\tEntropy+p)\vel) = - (\partial_c \mathcal{L}) R_c.
\end{equation}
As $\partial_c \mathcal{L} = - \sigma \Sigma_0/5$, we choose $R_c = - c/\tau$ to sign the mathematical entropy production $\varsigma \le 0$ and model the dissipation with an exponential decrease of characteristic time $\tau>0$. The final system reads
\begin{equation}
    \label{eq:two_scale_model_osc_unif_diss}
    \begin{cases}
        \setlength{\arraycolsep}{0pt}
        \begin{array}{lll}
            \partial_t  m_k &\:+ \:\bnabla \bcdot  (m_k \vel)&\: = 0,
            \qquad
            k=1,2,1^d, \\
            \partial_t  \alpha_1^d &\:+\: \bnabla \bcdot  (\alpha_1^d \vel)& \: = 0,                                                             \\
            \partial_t \Sigma_0 &\:+\:\bnabla \bcdot(\Sigma_0\vel)&\: = 0,\\
            \partial_t (\Delta\Sigma) &\:+\: \bnabla \bcdot(\Delta\Sigma\vel)&\: =-\tau^{-1}\Delta\Sigma,\\
            \partial_t  (\SigG) &\:+\: \bnabla \bcdot  (\SigG \vel) &\: = 0,                          \\
            \partial_t (\Sigma\left<H\right>_0) &\:+\: \bnabla \bcdot(\Sigma\left<H\right>_0\vel)&\: = 0,\\
            \partial_t (\rho \vel)&\:+\: \bnabla \bcdot (\rho \vel \otimes \vel+p \boldsymbol{I})&\: = 0,
        \end{array}
    \end{cases}
\end{equation}
with $p:=p_1=p_2$. One can see that the component of the interface area density modelling the droplets' dynamics have now its own dynamics with an additional source term that makes it dissipate over time.
\subsection{Modelling a spray of synchronously oscillating droplets}
\label{sec:org_osc}
%
%
%
%
With the choice of geometric quantities of the previous section, the moments are not suited to obtain a macroscopic oscillation of the spray.
Therefore, we introduce now new geometric quantities with GeoMOM based on another surface-average operator, and the following NDF approximation
\begin{equation}
    \label{eq:quadrature_osc}
    n_{\xi}(\widehat{S}_0,\widehat{\chi}, \widehat{\dot{\chi}}) = \sum_{i} n_i \delta(\widehat{S}_0 - (S_0)_i)\delta(\widehat{\chi}-\chi_i)\delta(\widehat{\dot{\chi}}-\dot{\chi}_i).
\end{equation}
This corresponds to several populations of droplets which share the same size and oscillate synchronously.
\subsubsection{GeoMOM based on an oriented surface-average operator}
%
%
%
%
%
Similarly to Section \ref{sec:geomom}, we consider a surface $\mathcal{S}$ and its mapping $\mathcal{U}\subset \mathbb{R}^2$ onto $\mathcal{S}\subset\mathbb{R}^3$ such that $A(u,v)\:dudv$ is the infinitesimal surface element over $\mathcal{S}$.
We decompose the surface local area into two contributions, one related to a preferred direction $\boldsymbol{N}$ given by the large-scale dynamics as suggested by \eqref{hyp:orientation}.
In this sense, we decompose this infinitesimal surface element $A(u,v)\:dudv$ using its definition using tangential vectors $\boldsymbol{e}_u$ and $\boldsymbol{e}_v$ 
\begin{equation}
    A = \Vert \boldsymbol{e}_u \times \boldsymbol{e}_v \Vert,
    \qquad
    A_{\parallel} = \left| (\boldsymbol{e}_u \times \boldsymbol{e}_v) \bcdot \boldsymbol{N} \right|,
    \qquad
    A_{\perp} = A - A_{\parallel}.
\end{equation}
From them, we decompose the surface into parallel and perpendicular components $S_{\parallel}$ and $S_{\perp}$ such that
\begin{equation}
    S = \int_{\mathcal{U}}A(u,v)\:dudv
    =
    \int_{\mathcal{U}}A_{\parallel}(u,v)\:dudv
    +
    \int_{\mathcal{U}}A_{\perp}(u,v)\:dudv
    = S_{\parallel} + S_{\perp}.
\end{equation}
We also define new surface-average operators similarly to~\eqref{eq:average_op}
\begin{equation}
    \left<\:\cdot\:\right>_{\parallel}:=\frac{1}{S_{\parallel}}\int_{\mathcal{U}}(\cdot)\: A_{\parallel}(u,v)\:dudv,
  \qquad
  \left<\:\cdot\:\right>_{\perp}:=\frac{1}{S_{\perp}}\int_{\mathcal{U}}(\cdot)\: A_{\perp}(u,v)\:dudv,
\end{equation}
and the surface-average operators $\widetilde{(\cdot)}^{\parallel}$, $\widetilde{(\cdot)}^{\perp}$ for one closed inclusion.
Considering now the oscillatory motion~\eqref{eq:small_scale_oscillator}, we
have the following dynamics for the geometric quantities of one droplet (see details in appendix \ref{app:diff_geom})
\begin{equation}
  \label{eq:def_geo_quant_oscil_paral_perp}
  \begin{aligned}
    S_{\parallel} = \frac{1}{2}S_0 - \frac{1}{\sqrt{4\pi}} \sqrt{\frac{5\pi}{2}}S_0\chi,
    \qquad
    &S_{\parallel}\widetilde{H}^{\parallel} = \frac{1}{2}\sqrt{4\pi}\sqrt{S_0} - \frac{1}{2}\sqrt{\frac{5\pi}{2}}\sqrt{S_0}\chi,\\
    S_{\perp} = \frac{1}{2}S_0 + \frac{1}{\sqrt{4\pi}} \sqrt{\frac{5\pi}{2}}S_0\chi,
    \qquad
    &S_{\perp}\widetilde{H}^{\perp} = \frac{1}{2}\sqrt{4\pi}\sqrt{S_0} + \frac{1}{2}\sqrt{\frac{5\pi}{2}}\sqrt{S_0}\chi.
  \end{aligned}
\end{equation}
Defining the oriented interface area densities $\Sigma_{\perp}$ and $\Sigma_{\parallel}$ within the mixture, we remark that $\Sigma = \Sigma_{\perp} + \Sigma_{\parallel}$ and $\Sigma\left<\:\cdot\:\right> = \Sigma_{\perp}\left<\:\cdot\:\right>_{\perp} + \Sigma_{\parallel}\left<\:\cdot\:\right>_{\parallel}$. Integrating~\eqref{eq:def_geo_quant_oscil_paral_perp} against $n_{\xi}$ provides
\begin{equation}
    \label{eq:rel_mom_geom_osc_sync}
  \begin{aligned}
    \Sigma_{\parallel} = \frac{1}{2}M^{\xi}_{1,0,0} - \frac{1}{\sqrt{4\pi}} \sqrt{\frac{5\pi}{2}}M^{\xi}_{1,1,0},
    \quad
    &(\Sigma\left<H\right>)_{\parallel} = \frac{1}{2}\sqrt{4\pi}M^{\xi}_{1/2,0,0}-\frac{1}{2}\sqrt{\frac{5\pi}{2}}M^{\xi}_{1/2,1,0},\\
    \Sigma_{\perp} = \frac{1}{2}M^{\xi}_{1,0,0} + \frac{1}{\sqrt{4\pi}} \sqrt{\frac{5\pi}{2}}M^{\xi}_{1,1,0},
    \quad
    &(\Sigma\left<H\right>)_{\perp} = \frac{1}{2}\sqrt{4\pi}M^{\xi}_{1/2,0,0}+\frac{1}{2}\sqrt{\frac{5\pi}{2}}M^{\xi}_{1/2,1,0}.
  \end{aligned}
\end{equation}
Remark that both the parallel and perpendicular geometric variables represent the same information for the spray as they are related to the same moments of the NDF.
Consequently, we here retain the perpendicular components only, and we extract the part dedicated to the dynamics by defining
\begin{equation}
    \label{eq:rel_mom_geom_osc_sync_decomp}
    \begin{aligned}
    \SigPerpEq&:= \tfrac{1}{2}M^{\xi}_{1,0,0},
    &\DSigPerp&:= \frac{1}{\sqrt{4\pi}} \sqrt{\frac{5\pi}{2}}M^{\xi}_{1,1,0}, \\
    \SigHPerpEq&:=\tfrac{1}{2}\sqrt{4\pi}M^{\xi}_{1/2,0,0},
    &\DSigHPerp&:=\frac{1}{2}\sqrt{\frac{5\pi}{2}}M^{\xi}_{1/2,1,0}.
    \end{aligned}
\end{equation}
We intentionally factored some indexes in the two last geometric quantities to lighten the notations.
This splitting allows us to retain more information on the NDF in our model by including two moments rather than the sum of two moments.
We know from~\eqref{eq:dyn_geom_var_S} that $\Sigma_{\perp,0}$ and $\SigHPerpEq$ are conserved, while we integrate~\eqref{eq:WBE_poly_vib} to obtain the dynamics of the two other geometric quantities
\begin{equation}
  \label{eq:dyn_geom_var_S_oriented}
  \begin{cases}
    \setlength{\arraycolsep}{0pt}
    \begin{array}{lll}
      \partial_t \DSigPerp &\:+ \: \bnabla \bcdot \left( \DSigPerp \vel\right) &\:= \frac{1}{\sqrt{4\pi}} \sqrt{\frac{5\pi}{2}} M^{\xi}_{1,0,1},\\
      \partial_t \DSigHPerp &\:+ \: \bnabla \bcdot \left( \DSigHPerp \vel\right) &\:= \frac{1}{2}\sqrt{\frac{5\pi}{2}} M^{\xi}_{1/2,0,1}.
    \end{array}
  \end{cases}
\end{equation}
With the moments considered up to now, we cannot model a macroscopic oscillatory spray as we lack some information regarding the distribution of $\widehat{\dot{\chi}}$ in our model.
We propose then to add two new geometric quantities that correspond to the unclosed moments of \eqref{eq:dyn_geom_var_S_oriented}
\begin{equation}
    \begin{aligned}
        \DtSigPerp
        &:= \Sigma_{\perp}\left<\frac{\partial_t A_{\perp}}{A_{\perp}}\right>_{\perp}
        &&= \frac{1}{\sqrt{4\pi}} \sqrt{\frac{5\pi}{2}} M^{\xi}_{1,0,1},\\
        \DtSigHPerp
        &:= \Sigma_{\perp}\left<\partial_t H + H \frac{\partial_t A_{\perp}}{A_{\perp}}\right>_{\perp}
        &&= \frac{1}{2}\sqrt{\frac{5\pi}{2}} M^{\xi}_{1/2,0,1}.
    \end{aligned}
\end{equation}
The above definitions with the oriented surface-average operators show that these quantities are also well-defined regardless of the flow regime.
Finally, we consider the following eight geometric quantities to describe the spray of oscillating droplets: $\alpha_1^d$, $\SigPerpEq$, $\DSigPerp$, $\DtSigPerp$, $\SigHPerpEq$, $\DSigHPerp$, $\DtSigHPerp$, $\SigG$.
\subsubsection{Amplitude-based closure}
With the eight moments given by the eight corresponding geometric quantities, we propose to look for a two-point quadrature which corresponds to two population of droplets as defined in \eqref{eq:quadrature_osc},
\begin{equation}
    \label{eq:quadrature_osc_two_pop}
    n_{\xi}(\widehat{S}_0,\widehat{\chi}, \widehat{\dot{\chi}}) = \sum_{i=1,2} n_i \delta(\widehat{S}_0 - (S_0)_i)\delta(\widehat{\chi}-\chi_i)\delta(\widehat{\dot{\chi}}-\dot{\chi}_i),
\end{equation}
where $n_i$ are the weights or numbers of droplets that share the same abscissas $\chi_i$ and $\dot{\chi}_i$.
The quadrature above admits a unique solution (excluding symmetry) under some realizability conditions that ensure that $\SigG$, $\SigHPerpEq$, $\Sigma_{\perp,0}$ and $\alpha_1^d$ are positive with additional geometric constraints (see appendix~\ref{app:polydisperse-closure}).
Remark that $n_i$ and $(S_0)_i$ depend only on $\SigG$, $\SigHPerpEq$, $\SigPerpEq$, $\VFracd$ and, for $(i,j)\in\{(1,2),(2,1)\}$,
\begin{equation}
    \label{eq:polydisperse_quadrature}
    \chi_i =\sqrt{\frac{2}{5\pi}}\frac{\sqrt{4\pi}\DSigPerp-2\sqrt{(S_{0})_j}\DSigHPerp}{n_i\left((S_{0})_i-\sqrt{(S_{0})_i(S_{0})_j}\right)},
    \quad
    \dot{\chi}_i =\sqrt{\frac{2}{5\pi}}\frac{\DtSigPerp-2\sqrt{(S_{0})_j}\DtSigHPerp}{n_i\left((S_{0})_i-\sqrt{(S_{0})_i(S_{0})_j}\right)}.
\end{equation}
We easily obtain that $\alpha_1^d$, $\SigG$, $\SigHPerpEq$ and $\SigPerpEq$ are conserved such that, together with \eqref{eq:dyn_geom_var_S_oriented}, it implies
\begin{equation}
    \label{eq:abscissa_dynamics}
    \partial_t n_i + \bnabla \bcdot  (n_i\vel)= 0,
    \quad
    D_t (S_{0})_i = 0,
    \quad
    D_t\chi_i = \dot{\chi}_i,
    \quad
    i=1, 2.
\end{equation}
This dynamics is expected as the two population of oscillating droplets are advected by the flow at velocity $\vel$.
\subsubsection{Two-scale model with the small-scale spray model of synchronous droplets}
Denote $\nu := \rho_1^d/(4(4\pi)^{3/2})$ and $\gamma := 2\sigma$ such the kinetic and potential energies of the spray of oscillating droplets given in \eqref{eq:osc_energies} write
\begin{equation}
  E^{kin,d} = \frac{1}{2} \nu M^{\xi}_{5/2,0,2}, \qquad E^{pot,d} = \sigma M^{\xi}_{1,0,0} + \frac{1}{2}\gamma M^{\xi}_{1,2,0},
\end{equation}
and $\tilde{\omega}^2 = \gamma/\nu$.
The closure~\eqref{eq:quadrature_osc_two_pop} then yields
\begin{equation}
    \label{eq:vibrating_energies_poly}
    E^{kin,d} = \sum_{i=1,2} \frac{1}{2}\nu n_i (S_0)_i^{5/2}\dot{\chi}_i^2,
    \qquad
    E^{pot,d} = \sum_{i=1,2} \sigma n_i (S_0)_i + \frac{1}{2}\gamma n_i (S_0)_i \chi_i^2
\end{equation}
We extend the two-scale Lagrangian \eqref{eq:lagrangian_two_scale} by adding the energies above to account for the small-scale oscillation
\begin{equation}
    \label{eq:lagrangian_oscill}
    \begin{aligned}
    \lag =& 
    \lag_1\left(\alpha_1, m_1, \vel\right)
    +\lag_2\left(\alpha_2, m_2, \vel\right)
    +\lag_{1}^d\left(m_1^d, \rho_1^d, \vel\right)\\
    &+\lag^{vib}_1\left(n_1, (S_0)_1, \chi_1, \dot{\chi}_1\right)
    +\lag^{vib}_2\left(n_2, (S_0)_2, \chi_2, \dot{\chi}_1\right),
  \end{aligned}
\end{equation}
where $\lag^{vib}_i$ is defined using vibrating energies of (\ref{eq:vibrating_energies_poly})
\begin{equation}
    \label{eq:lagrangian_osc_small_scale}
    \lag^{vib}_i
    =
    \frac{1}{2}\nu n_i (S_0)_i^{5/2}\dot{\chi}_i^2
    - \sigma n_i (S_0)_i
    - \frac{1}{2}\gamma n_i (S_0)_i \chi_i^2.
\end{equation}
Remark that the kinetic energy is here positively signed as it is a quadratic form of $\dot{\chi}_1$ and $\dot{\chi}_2$ with their associated momentum equations after Hamilton's SAP.
We provided an expression using the quadrature's abscissas rather than the geometric quantities for computational convenience, but the dynamics of the geometric quantities is equivalently obtained using the quadrature expression of appendix \ref{app:polydisperse-closure}.
In addition to $\alpha_1$, the quantities $\chi_1$ and $\chi_2$ are also free variables which results in the following system with two additional momentum equations for each population of droplets of same size in the spray (see appendix \ref{app:SAP_osc})
\begin{equation}
    \label{eq:system_poly_osc}
    \begin{cases}
        \setlength{\arraycolsep}{1pt}
        \begin{array}{llll}
            \partial_t m_k &+ \bnabla \bcdot (m_k \vel)&=0,   &\qquad k=1,2,1^d,\\
            \partial_t n_i &+ \bnabla \bcdot (n_i \vel)&=0,   &\qquad i=1,2,\\
            \partial_t (n_i(S_0)_i) &+ \bnabla \bcdot (n_i(S_0)_i \vel) &=0,   &\qquad i=1,2,\\
            \partial_t(n_i\dot{\chi}_i)&+ \bnabla \bcdot (n_i\dot{\chi}_i \vel)&=-\omega_i^2 n_i\chi_i, &\qquad i=1,2, \\
            \partial_t(n_i\chi_i)&+ \bnabla \bcdot (n_i\chi_i \vel) &=n_i\dot{\chi}_i, &\qquad i=1,2,\\
            \partial_t (\rho \vel) &+ \bnabla \bcdot (\rho \vel \otimes \vel+p)&=0, &                    
        \end{array}
    \end{cases}
\end{equation}
with $p:=p_1=p_2$ and $\omega_i^2 = \tilde{\omega}^2 (S_0)_i^{-3/2}$. Combining the equations at the fifth and sixth lines of the system above, one can recognize the equations of harmonic oscillators advected along the streamlines
\begin{equation}
    D_t(D_t \chi_i) + \omega_i^2 \chi_i = 0,
    \qquad
    i=1,2.
\end{equation}
Similarly to the previous models, this system admits an additional conservation equation on the total energy $\tEntropy$ defined hereafter 
\begin{equation}
    \partial_t \tEntropy + \bnabla \bcdot ((\tEntropy+p)\vel)=0,
    \qquad
    \tEntropy = \sum_{k=1,2,1^d} \frac{1}{2}m_k\vert\vel\vert^2 + \sum_{i=1,2} \frac{1}{2}\nu n_i (S_0)_i^{5/2}\dot{\chi}_i^2 -\lag.
\end{equation}
Once again, we could consider a pressure relaxation model, but we would like to focus again on the dissipation associated to the oscillation process, and we introduce source terms $R_{\chi_i}$ in the new momentum equations
\begin{equation}
    \partial_t(n_i\dot{\chi}_i)
    + \bnabla \bcdot (n_i\dot{\chi}_i \vel)
    =-\omega_i^2 n_i\chi_i
    + R_{\chi_i}
    \qquad
    i=1,2.
\end{equation}
This source term provides the following mathematical entropy production
\begin{equation}
    \varsigma := \partial_t \tEntropy + \bnabla \bcdot ((\tEntropy+p)\vel)
    =\sum_{i=1,2}\nu (S_0)_i^{5/2}R_{\chi_i}D_t\chi_i.
\end{equation}
As the closure \eqref{eq:quadrature_osc_two_pop} groups droplets by size, we can now model the first-order size-dependent damping term of the viscous droplet in a light carrier phase \citep{prosperetti_viscous_1977,plumacher_non-linear_2020}.
Then, we expect each population of oscillators to be damped following
\begin{equation}
    \label{eq:viscous_droplet_model}
    D_{t}(D_t(\chi_i))+\omega_i^2 \chi_i=-\beta_i D_t\chi_i
    \quad
    \iff
    \quad
    D_{t}\dot{\chi}_i+\omega_i^2 \chi_i=-\beta_i \dot{\chi}_i,
\end{equation}
with $\beta_i=4\pi\nu_{vis}/(S_0)_i>0$ with $\nu_{vis}$ the liquid kinematic viscosity.
One can then choose $R_{\chi_i}=-n_i\beta_i\dot{\chi}_i$ and recover both the above dissipation process for both populations of droplets and a signed production of mathematical entropy 
\begin{equation}
    \varsigma
    =-\beta_1n_1\nu(S_0)_1^{5/2}\dot{\chi}_1^2
    -\beta_2n_2\nu(S_0)_2^{5/2}\dot{\chi}_2^2
    \le 0.
\end{equation}
It leads to the following dissipative model
\begin{equation}
    \label{eq:system_poly_osc_diss}
    \begin{cases}
        \setlength{\arraycolsep}{1pt}
        \begin{array}{llll}
            \partial_t m_k &+ \bnabla \bcdot (m_k \vel)&=0,   &\qquad k=1,2,1^d,\\
            \partial_t n_i &+ \bnabla \bcdot (n_i \vel)&=0,   &\qquad i=1,2,\\
            \partial_t (n_i(S_0)_i) &+ \bnabla \bcdot (n_i(S_0)_i \vel) &=0,   &\qquad i=1,2,\\
            \partial_t(n_i\dot{\chi}_i)&+ \bnabla \bcdot (n_i\dot{\chi}_i \vel)&=-\omega_i^2 n_i\chi_i-\beta_in_i\dot{\chi}_i, &\qquad i=1,2, \\
            \partial_t(n_i\chi_i)&+ \bnabla \bcdot (n_i\chi_i \vel) &=n_i\dot{\chi}_i, &\qquad i=1,2,\\
            \partial_t (\rho \vel) &+ \bnabla \bcdot (\rho \vel \otimes \vel+p \boldsymbol{I})&=0, &                   
        \end{array}
    \end{cases}
\end{equation}
with $p:=p_1=p_2$.
This model shows a macroscopic synchronous oscillation through the two momentum equations on $\dot{\chi}_1$ and  $\dot{\chi}_2$ and includes a physics-based dissipation rate.
The system is here written using the weights and abscissas, but it can also be written using the geometric quantities of the model. The interface area density dynamics is specifically discussed in the next section.
%
%
%
%
%
%
%

%% file: 4_geometry_dyn.tex
\section{Dynamics of the small-scale interface area density}
\label{sec:geometry_dynamics}
To perceive the importance of the choice of variables and modelling choices made in the previous sections, in this last section we investigate their impact on the  closure of the evolution equation for the interface area density.
Such an equation is usually obtained from an averaging process and, whereas several terms are classically identified, their closure is most of the time out of reach except in very simplified configurations. 
We show how the proper choice of variables for both the static minimal surface spherical case, but also the dynamical case, allows a clear-cut strategy to close this important evolution equation and rethink the set of variables we should use for a unified model.

\subsection{Geometry of the disperse regime through the classical averaging approach}
First, the time evolution of the geometry for a generic two-phase flow is derived with an averaging process \citep{drew_theory_1999, lhuillier_evolution_2004, morel_mathematical_2015}.
This approach relies on the kinematics of the interface where the phases $k=1,2$ are located using a phase indicator function $\boldsymbol{X}_k$ and the interface is identified through its derivative $\bnabla\boldsymbol{X}_k$ in the sense of generalized function.
Denoting $\vel_I$ the interface velocity, the kinematics of the interface reads 
\begin{equation}
    \partial_t \boldsymbol{X}_k + \vel_I \bcdot \bnabla \boldsymbol{X}_k = 0.
\end{equation}
The ensemble average operator $\left<\:\cdot\:\right>_E$ is introduced such that the volume fraction and the interface area density are defined by $\alpha_k~=~\left<\boldsymbol{X}\right>_E$ and $\Sigma=\left<\delta_I\right>_E$ where $\delta_I=-\boldsymbol{n}\bnabla\boldsymbol{X}_k$ is the interface generalized function. 
Then, the evolution of the interface area density is derived in \cite{lhuillier_evolution_2004} as the trace of the interface tensor $\delta_I \boldsymbol{n}\otimes\boldsymbol{n}$. It eventually yields
\begin{equation}
    \begin{cases}
        \partial_t\alpha + \left<\vel_I\bcdot\bnabla \boldsymbol{X}\right>_E=0,\\
        \partial_t \Sigma + \bnabla \bcdot \left<\delta_I\vel_I\right>_E
        =\left<\boldsymbol{I}-\boldsymbol{n}\otimes\boldsymbol{n}:\bnabla\vel_I \delta_I\right>_E.
    \end{cases}
\end{equation}
Such an averaging approach is compatible with the kinetic-based small-scale models of this work in the statistical sense.
Similarly, the characteristics (size, oscillation amplitude, ...) are following the probabilistic law given by the NDF.
Focusing on the small-scale disperse regime while assuming \eqref{hyp:ls_shared_vel}, we have $\alpha=\alpha_1^d$, and we decompose $\vel_I = \vel + v_n \boldsymbol{n}$.
Moreover, the symmetry of either spherical inclusions \eqref{hyp:spheres} or oscillation motion \eqref{hyp:osc_sec_harm} gives $\left<v_n\boldsymbol{n}\delta_I\right>_E=0$.
The averaged equations then become 
\begin{equation}
    \label{eq:interface_kinetmatics}
    \begin{cases}
        D_t \alpha_1^d = -\left<v_n\boldsymbol{n}\bcdot\bnabla \boldsymbol{X}\right>_E,\\
        \partial_t \Sigma + \bnabla \bcdot (\Sigma \vel)=
        \frac{2}{3}\Sigma\bnabla\bcdot\vel
        -\left<\boldsymbol{q}:\bnabla\vel\right>_E
        +\frac{2}{3}\left<\bnabla\bcdot(v_n\boldsymbol{n})\delta_I\right>_E
        -\left<\boldsymbol{q}:\bnabla(v_n\boldsymbol{n})\right>_E,
    \end{cases}
\end{equation}
where $\boldsymbol{q} = \left<(\boldsymbol{n}\otimes \boldsymbol{n}-\boldsymbol{I}/3)\delta_I\right>_E$ is the anisotropic tensor.
In this last equation, one can identify the different contributions to the interface area density evolution: (from left to right of the right-hand side) the large-scale isotropic and anisotropic terms, the small-scale isotropic and anisotropic terms.
\subsection{Comparison with the small-scale models}
From the different small-scale models in this work, as well as proper choice of variables, we show that the interface area density equation of evolution can be thoroughly closed.
We compare it to the unclosed kinematic set of equations \eqref{eq:interface_kinetmatics} and identify the various contributions.
The question of the choice of fundamental variables to describe the dynamics of the interface is eventually discussed.
\subsubsection{Spray of compressible spherical inclusions}
Before considering an incompressible small scale dedicated to the description of droplets, we focus on the case of compressible inclusions such as bubbles, which has been partially treated in Section \ref{sec:polydisperse}.
We recast the unclosed dynamics for $\Sigma$ obtained in \eqref{eq:geom_dynamics_compressible} into
\begin{equation}
    \partial_t\Sigma + \bnabla \bcdot (\Sigma \vel) = \frac{2}{3}\Sigma \bnabla \bcdot \vel + \frac{2}{3} \Sigma \frac{D_t \alpha_1^d}{\alpha_1^d}.
\end{equation}
Even though the previous equation is unclosed as the dynamics of $\alpha_1^d$ is not specified, one can still identify the isotropic contribution of the large scale of \eqref{eq:interface_kinetmatics}.
As we consider spherical shapes, $\boldsymbol{q}=\boldsymbol{0}$ and there are no anisotropic contributions. Therefore, the small-scale isotropic term is
\begin{equation}
    \left<\bnabla\bcdot(v_n\boldsymbol{n})\delta_I\right>_E = \Sigma \frac{D_t \alpha_1^d}{\alpha_1^d}.
\end{equation}
A closed model is thus obtained with the dynamics of $\alpha_1^d$ as in the next case dealing with incompressible inclusions.

\subsubsection{Spray of incompressible spherical droplets}

The incompressible case has been treated in Section \ref{sec:polydisperse} after assuming $D_t \rho_1^d = 0$. We have then the following conservation equations
\begin{equation}
    \begin{cases}
        \setlength{\arraycolsep}{0pt}
        \begin{array}{lll}
            \partial_t \alpha_1^d &+ \bnabla \bcdot (\alpha_1^d \vel) &= 0,\\
            \partial_t \Sigma &+ \bnabla \bcdot (\Sigma \vel) &= 0.
        \end{array}
    \end{cases}
\end{equation}
The anisotropic tensor $\boldsymbol{q}$ is still nil and provides trivial anisotropic closures, while the isotropic closures read
\begin{equation}
    \left<v_n\boldsymbol{n}\bcdot\bnabla \boldsymbol{X}\right>_E
    =
    \alpha_1^d \bnabla\bcdot\vel,
    \qquad
    \left<\bnabla\bcdot(v_n\boldsymbol{n})\delta_I\right>_E
    =
    -\Sigma\bnabla\bcdot\vel.
\end{equation}
This shows that the small-scale isotropic contributions balance the large-scale ones to maintain the incompressibility of the small-scale.
\subsubsection{Two-scale model with the small-scale spray model of asynchronous droplets}
This case has been treated in Section \ref{sec:disorg_osc}, where the oscillatory dynamics of the incompressible droplets is formulated in a decomposed form of the small-scale interface area density following $\Sigma = \Sigma_0 + \Delta\Sigma$. This results in the following set of equations
\begin{equation}
    \begin{cases}
        \setlength{\arraycolsep}{0pt}
        \begin{array}{lll}
            \partial_t \alpha_1^d &+ \bnabla \bcdot (\alpha_1^d \vel) &= 0,\\
            \partial_t \Sigma &+ \bnabla \bcdot (\Sigma \vel) 
            &= - \frac{\Delta\Sigma}{\tau}.
        \end{array}
    \end{cases}
\end{equation}
Following the assumption (\ref{hyp:orientation}) made on the orientation of the oscillating droplets, one can consider the right-hand side term as either isotropic or anisotropic
\begin{equation}
    \begin{cases}
        \setlength{\arraycolsep}{0pt}
        \begin{array}{lll}
        \left<v_n\boldsymbol{n}\bcdot\bnabla \boldsymbol{X}\right>_E
        &=&
        \alpha_1^d \bnabla\bcdot\vel,\\
        \left<\boldsymbol{q}:\bnabla\vel\right>_E &=& 0,
        \end{array}\\
        \frac{2}{3}\left<\bnabla\bcdot(v_n\boldsymbol{n})\delta_I\right>_E
        -\left<\boldsymbol{q}:\bnabla(v_n\boldsymbol{n})\right>_E
        =
        -\frac{2}{3}\Sigma\bnabla\bcdot\vel
        -\frac{\Delta\Sigma}{\tau}.
    \end{cases}
\end{equation}
The excess of interface area density $\Delta\Sigma$ is exponentially decreasing towards $0$. One could have make other dissipative choices such as $R_c \propto - c^2$ to get a decreasing rate similar to the destruction source term as in \cite{vallet_modelisation_1999,anez_eulerianlagrangian_2019} which is quadratic in $\Sigma$. The creation terms present in these works result in an equilibrium value for $\Delta\Sigma$.
Such terms cannot be recovered here as it would require a source term, for instance to model a small-scale turbulent energy which is not here accounted for. This is the subject of a current work in progress. 
\subsubsection{Two-scale model with the small-scale spray model of synchronous droplets}
This case has been studied in Section \ref{sec:org_osc}, where incompressible droplets oscillate synchronously for a given size.
The dynamics has been derived respectively for each size of droplets using the weights and abscissas rather than the geometric quantities.
Without losing the properties of the polydisperse case, we propose here to focus on the monodisperse case detailed in appendix \ref{app:monodisperse}.
It allows a more compact equation of evolution for the small-scale interfacial area density $\Sigma = M_{1,0,0}^{\xi} + M_{1,2,0}^{\xi}$
\begin{equation}
    \label{eq:sigma_eq_mono}
    \partial_t \Sigma + \bnabla \bcdot (\Sigma \vel)  = \frac{16}{5}
    \frac{\DSigPerp\DtSigPerp}{(n_1^d)^{1/3}(6\sqrt{\pi}\alpha_1^d)^{2/3}}.
\end{equation}
Similarly to the previous cases, the incompressibility condition balances the isotropic terms to conserve $\alpha_1^d$.
Moreover, the right-hand side of \eqref{eq:sigma_eq_mono} only accounts for the small-scale anisotropic motion such that
\begin{equation}
    \begin{cases}
        \begin{array}{lll}
        \left<v_n\boldsymbol{n}\bcdot\bnabla \boldsymbol{X}\right>_E
        &=&
        \alpha_1^d \bnabla\bcdot\vel,\\
        \left<\bnabla\bcdot(v_n\boldsymbol{n})\delta_I\right>_E
        &=&
        -\Sigma\bnabla\bcdot\vel,\\
        \left<\boldsymbol{q}:\bnabla\vel\right>_E&=&0,\\
        \left<\boldsymbol{q}:\bnabla(v_n\boldsymbol{n})\right>_E &=& -\frac{2}{(3\sqrt{4\pi})^{2/3}} \frac{\DSigPerp\DtSigPerp}{(n_1^d)^{1/3}(\alpha_1^d)^{2/3}}.
        \end{array}
    \end{cases}
\end{equation}
This last identification of closures concludes our approach where the dynamics is first assessed, and the geometry kinematics is obtained secondly.
Our approach shows that for this last case, the dynamics cannot be simply expressed relying on the usual quantities $\alpha_1^d$ and $\Sigma$. 
The SAP modelling strategy allows to identify the proper set of variables and conservation equations in the dynamical case, and to obtain a closed conservation equation for the interface area density in particular, an important building block for further modelling such as evaporation or heat transfer.

%% file: 5_conclusion.tex
\section{Conclusion}
In this work, we have proposed a novel framework to derive two-scale reduced-order models based on Hamilton's SAP as well as a set of geometric variables leading to the premises of a unified model for both disperse and separated phases two-phase flows, potentially including several dissipation phenomena.
A hierarchy of small-scale models involving a variety of physical phenomena can be described within the framework.

The model is compatible in the two limits with classical models of the literature for disperse and separated two-phase flows and possesses essential properties for a proper mathematical framework, that is  hyperbolicity and signed mathematical entropy evolution. 

To deal with interface dynamics at small scale, we rely on GeoMOM; geometric variables are defined, which can be interpreted both as moments of a kinetic description at small scale, or as surface averaged quantities, which are also defined in other regimes without any assumption on the geometry of the interface.
This makes the link with the mixed zone and open new perspectives since we are able to tackle interface dynamics at small scale and reach a closed surface area density evolution equation for all the proposed small-scale models.
It sheds some light on the fact that the natural variables to tackle the interface dynamics are not necessarily the interface area density and volume fractions but more intrinsic geometric quantities, the dynamics of which allow to recover the interface area density evolution. 

Two issues have been left aside on purpose in the design of the paper for the sake of clarity of the exposition:
1- the existence of multiple velocities for the large-scale phases and for the disperse small-scale,
2- the transfer of mass from large-scale to small-scale. The key issue is to include the related physics within the proposed framework.
The first part is currently under investigation and relates to another version of the SAP with multiple velocities, while the second requires to include the capillarity at large scale and to interpret the transfer of scales as a local dissipative phenomenon; it is also the subject of a complementary piece of work \citep{loison_two-scale_2023}.
Eventually, numerical methods also have to be designed in order to resolve properly the proposed models in the line of GeoMOM and preserve realizability \citep{ait-ameur_simulation_2023}.
One last key issue is related to the treatment of small-scale agitation/turbulence of the large-scale phase, as mentioned in Section~\ref{sec:kinetic-model-oscill}, which would couple this velocity fluctuations to the droplet oscillations and provide sources terms in the kinetic equation.
This is also the subject of our current investigation.
This non-exhaustive list of extensions tends to assess the versatility of the proposed framework for the design of physically relevant and mathematically well-designed two-scale models for interfacial two-phase flows.

%% file: SAP_comp_incomp.tex
\section{Condition of small-scale incompressibility in the disperse regime}
\label{app:incomp}
We are interested in the compressibility caused by sound propagation at large scale.
Consider then the two-scale mixture defined in Section \ref{sec:two-scale} with two large-scale liquid and gaseous phases and one small-scale liquid phase that share the same pressure $P$.
Denote $\kappa_T^k := \rho_k^{-1}(\partial \rho_k/\partial P)_T$ the isothermal compressibility, we measure the mixture compressibility using the mass conservation of each phase with
\begin{equation}
    \bnabla\bcdot\vel = \sum_{k=1,2,1^d}\alpha_k \bnabla\bcdot\vel
    =-\sum_{k=1,2,1^d}\alpha_k\frac{D_t\rho_k}{\rho_k}
    =-\sum_{k=1,2,1^d}\alpha_k\kappa_T^k D_t P.
\end{equation}
The liquid phase is much less compressible than the gaseous ones $\kappa_T^1,\kappa_T^{1,d}\ll\kappa_T^2$.
Moreover, the small-scale liquid phase only occupies a small amount of the mixture volume $\alpha_1^d~\ll~\alpha_1,~\alpha_2$. 
Then, the contributions to the overall compressibility of the mixture are ranked like
\begin{equation}
    \vert\alpha_2\kappa_T^2 D_t P\vert
    \gg
    \vert\alpha_1\kappa_T^1 D_t P\vert
    \gg
    \vert\alpha_1^d\kappa_T^{1,d} D_t P\vert.
\end{equation}
The small-scale liquid phase can then be assumed incompressible in comparison to both large-scale phases.

%% file: SAP_two_scale.tex
\section{Stationary Action Principle}
\label{app:SAP}

\subsection{Derivation of the conservative dynamics}

Consider the Lagrangian defined in \eqref{eq:lagrangian_two_scale},
\begin{equation}
    \lag = \lag_1(\alpha_1, m_1, \vel) + \lag_2(\alpha_2, m_2, \vel) + \lag_1^d(m_1^d, \rho_1^d, \vel).
\end{equation}
We minimize the action of the whole mixture defined by
\begin{equation}
    \mathcal{A} = \int_{\Omega} \lag d\boldsymbol{x}dt,
\end{equation}
where $\boldsymbol{x}$ is the position in Eulerian coordinates and $\Omega:=\Omega_{\boldsymbol{x}}\times[0,T]$ is the Eulerian space-time domain.
This minimization is performed over a family of trajectories defined by Lagrangian mappings $\boldsymbol{\phi}^{\lambda}(\boldsymbol{X},t, \lambda)$ parametrized by $\lambda$ which lies in the vicinity of $0$ and $\boldsymbol{X}=\boldsymbol{\phi}^{-1}(\boldsymbol{x},t)$ the Lagrangian coordinates. 
We similarly introduce families of Eulerian fields $\alpha_{1}^{\lambda}(\boldsymbol{x},t, \lambda)$, $b_{c}^{\lambda}(\boldsymbol{x},t, \lambda)$, $b_{a}^{\lambda}(\boldsymbol{x},t, \lambda)$ for the volume fraction, the conserved variables and the advected ones.
We assume that these families of Lagrangian mappings and Eulerian fields satisfy the following conditions.
\begin{itemize}
    \item The mapping and Eulerian fields of the solution is included in the families for $\lambda=0$ \textit{i.e.} for all $(\boldsymbol{x},t)\in \Omega$,
    \begin{equation}
            \alpha_{1}^{\lambda}(\boldsymbol{x},t,\lambda=0)=\alpha_1(\boldsymbol{x},t),
            \quad
            b_{c}^{\lambda}(\boldsymbol{x},t,\lambda=0)=b_{c}(\boldsymbol{x},t),\quad
            b_{a}^{\lambda}(\boldsymbol{x},t,\lambda=0)=b_{a}(\boldsymbol{x},t).
    \end{equation}
    \item All the mappings and Eulerian fields preserve the constraints \textit{i.e.} for all $(\boldsymbol{x},t)\in \Omega$,
    \begin{equation}
        \partial_t b_c^{\lambda} + \bnabla\bcdot(b_c^{\lambda}\vel)=0,
        \qquad
        \partial_t b_a^{\lambda} + \vel\bcdot\bnabla b_a^{\lambda}=0.
    \end{equation}
    \item All the mappings and Eulerian fields preserve the values at boundaries of the space-time domain \textit{i.e.} for all $(\boldsymbol{x},t)\in \partial\Omega$,
    \begin{equation}
        \alpha_{1}^{\lambda}(\boldsymbol{x},t,\lambda)=\alpha_1(\boldsymbol{x},t),
        \quad
        b_{c}^{\lambda}(\boldsymbol{x},t,\lambda)=b_{c}(\boldsymbol{x},t),
        \quad
        b_{a}^{\lambda}(\boldsymbol{x},t,\lambda)=b_{a}(\boldsymbol{x},t).
    \end{equation}
\end{itemize}
We define the following variations,
\begin{equation}
    \begin{aligned}
        \boldsymbol{\VarTraj}(\boldsymbol{x},t) := \left(\partial_{\lambda}\boldsymbol{\phi}^{\lambda}\right)_{\boldsymbol{X},t}(\boldsymbol{\phi}^{-1}(\boldsymbol{x},t),t, \lambda=0),
        \qquad
        \delta b_e(\boldsymbol{x},t) := \left(\partial_{\lambda} b^{\lambda}_e\right)_{\boldsymbol{x},t}(\boldsymbol{x},t,\lambda=0),
    \end{aligned}
\end{equation}
where $\boldsymbol{\VarTraj}$ is an infinitesimal Eulerian displacement and $\delta b_e$ the variation on any Eulerian field $b_e$.
With this variational operator, Hamilton's SAP simply writes
\begin{equation}
    \delta\mathcal{A}=0.
\end{equation}
Following \cite{fosdick_kinks_1991, gavrilyuk_hyperbolic_1998, gouin_hamiltons_1999, gavrilyuk_mathematical_2002, drui_small-scale_2019,gouin_introduction_2020,cordesse_diffuse_2020}, the variations of $b_c$, $b_a$ and $\vel$ are related to $\boldsymbol{\VarTraj}$ through relations
\begin{equation}
    \label{eq:variation_constraints}
    \delta b_c = - \bnabla \bcdot (b_c\boldsymbol{\VarTraj}),
    \qquad
    \delta b_a = -(\boldsymbol{\VarTraj}\bcdot\bnabla)b_a,
    \qquad
    \delta\vel = D_t \boldsymbol{\VarTraj}-(\boldsymbol{\VarTraj}\bcdot\bnabla)\vel.
\end{equation} 
For the Lagrangian under consideration, we have $b_c\in\{m_1, m_2, m_1^d, \alpha_1^d\}$ and advected quantities $b_a\in\{\rho_1^d\}$.
The variation of $\alpha_2$ is linked to both $\boldsymbol{\VarTraj}$ and $\delta\alpha_1$ through the volume occupation relation $\alpha_2=1-\alpha_1-\alpha_1^d$. 
With the summation on repeated indexes, we define the divergence  $\boldsymbol{A}$ by $\bnabla\bcdot\boldsymbol{A}=(\partial_{x_j}A_{ij})$ and the gradient of a vector $\bnabla\boldsymbol{b}=(\partial_{x_i}b_j)$.
We decompose the action variation in Eulerian coordinates with respect to each dependency
\begin{align}
    \delta\Action_{\alpha_1} & = \int_\Omega \partial_{\alpha_1}\lag_1 \: \delta\alpha_1,\\
    \delta\Action_{\alpha_2} & = \int_\Omega \partial_{\alpha_2}\lag_2 \: \delta\alpha_2
    =-\int_{\Omega}\alpha_1^d \bnabla (\partial_{\alpha_2}\lag_2)\bcdot\VarTraj
    -\int_{\Omega}\partial_{\alpha_2}\lag_2 \: \delta \alpha_1,\\
    \delta\Action_{m_k} & = \int_\Omega \partial_{m_k}\lag_k  \: \delta m_k = \int_\Omega m_k \bnabla (\partial_{m_k}\lag_k)\bcdot \VarTraj,\\
    \delta\Action_{\rho_1^d} & = \int_\Omega -\partial_{\rho_1^d}\lag_1^d \: \bnabla\rho_1^d \bcdot \VarTraj,\\
    \delta\Action_{\vel} & = \int_\Omega -\left(\partial_t\boldsymbol{K}+\bnabla\bcdot(\boldsymbol{K} \otimes \vel)+\boldsymbol{K}\bcdot\bnabla\vel\right)\bcdot\VarTraj,
\end{align}
where $\delta\alpha_1^d$ follows (\ref{eq:variation_constraints}), $\delta\alpha_2=-\delta\alpha_1-\delta\alpha_1^d$ and $\boldsymbol{K}=\partial_{\vel}\lag$.
Remark that we denoted here the matrix-vector product of a matrix $\boldsymbol{A}$ with a vector $\boldsymbol{b}$ by $\boldsymbol{b}\bcdot\boldsymbol{A}=(A_{ij}b_j)$.
Denoting $\lag_k^* = m_k(\partial_{m_k} \lag_k)-\lag_k$ and $\lag^* = \sum_k \lag_k^*$, then $\delta\Action$ can be expressed as a combination of variations $\boldsymbol{\VarTraj}$ and $\delta\alpha_1$,
\begin{equation}\label{eq:action_variation}
    \begin{aligned}
        \delta \Action =&-\int_{\Omega}\left(\partial_t\boldsymbol{K}+\bnabla\bcdot(\boldsymbol{K} \otimes \vel)+\boldsymbol{K}\bcdot\bnabla\vel+\alpha_1^d \bnabla (\partial_{\alpha_2}\lag_2)+\partial_{\rho_1^d}\lag_1^d\:\bnabla\rho_1^d \right.\\
        &\qquad\left.-m_1 \bnabla (\partial_{m_1}\lag_1)-m_2 \bnabla (\partial_{m_2}\lag_2)-m_1^d \bnabla (\partial_{m_1^d}\lag_1^d)\right)\bcdot\VarTraj\\
        &+\int_{\Omega}(\partial_{\alpha_1}\lag_1-\partial_{\alpha_2}\lag_2) \: \delta \alpha_1,\\
        =& -\int_{\Omega} (\partial_t \boldsymbol{K}+\bnabla\bcdot(\boldsymbol{K} \otimes \vel) - \bnabla( \lag^*-\alpha_1^d\partial_{\alpha_2}\lag_2)-(\partial_{\alpha_1}\lag_1-\partial_{\alpha_2}\lag_2)\bnabla\alpha_1)\bcdot \VarTraj\\
        & +\int_{\Omega}(\partial_{\alpha_1}\lag_1-\partial_{\alpha_2}\lag_2)\:\delta\alpha_1,
    \end{aligned}
\end{equation}
with $\boldsymbol{K}=\partial_{\vel} \lag$ and $\lag_k^* = m_k(\partial_{m_k} \lag_k)-\lag_k$.
The fundamental theorem of variations applied to (\ref{eq:action_variation}) gives
\begin{equation}
    \label{eq:null_variations}
    \delta \Action=0
    \Rightarrow
    \begin{cases}
        \partial_t\boldsymbol{K}+\bnabla\bcdot(\boldsymbol{K} \otimes \vel) - \bnabla( \lag^*-\alpha_1^d\partial_{\alpha_2}\lag_2) = 0, \\
        \partial_{\alpha_1}\lag_1-\partial_{\alpha_2}\lag_2 = 0.
    \end{cases}
\end{equation}
Developing the expression of the derivatives yields \eqref{eq:two_scale-incomp-cons}.
\subsection{Additional conservation equation}
Starting from \eqref{eq:two_scale-incomp-cons}, we now look for an additional conservation equation on $\tEntropy = \boldsymbol{K}\bcdot\vel - \lag$. We take the scalar product of the momentum equation with $\vel$ to get
\begin{equation}
    \label{eq:H_compute}
    \begin{aligned}
        0=\:&\partial_t(\boldsymbol{K}\bcdot\vel)-\boldsymbol{K}\partial_t \vel+\bnabla\bcdot((\boldsymbol{K}\bcdot\vel) \vel) -\boldsymbol{K}(\vel\bcdot\bnabla)\vel- \bnabla\bcdot(p\vel)-p\bnabla\bcdot\vel\\[5pt]
        =\:&\partial_t \tEntropy +\bnabla\bcdot((\tEntropy+p) \vel)
        + \partial_t \lag -\boldsymbol{K}D_t \vel+\bnabla\bcdot(\lag \vel) -p\bnabla\bcdot\vel\\[5pt]
        =\:&\partial_t \tEntropy +\bnabla\bcdot((\tEntropy+p) \vel)
        + D_t \lag -\boldsymbol{K}D_t \vel+ (\lag - p)\bnabla\bcdot\vel.
    \end{aligned}
\end{equation}
We develop the material time derivative and isolate the mass and incompressibility constraints such that \eqref{eq:H_compute} becomes
\begin{equation}
    \label{eq:H_compute_2}
    \begin{aligned}
        0=\:&\partial_t \tEntropy +\bnabla\bcdot((\tEntropy-(\lag^*-\alpha_1^d\partial_{\alpha_2}\lag_2)) \vel)
        + (\lag - p)\bnabla\bcdot\vel\\
        &\qquad + \partial_{\alpha_1}\lag_1 \: D_t \alpha_1+ \partial_{\alpha_2}\lag_2 \: D_t \alpha_2
        + \partial_{\rho_1^d}\lag_1 D_t \rho_1^d\\
        &\qquad + \partial_{m_1} \lag_1 D_t m_1
        + \partial_{m_2} \lag_2 D_t m_2
        + \partial_{m_1^d} \lag_1^d D_t m_1^d\\[5pt]
        =\:&\partial_t \tEntropy +\bnabla\bcdot((\tEntropy+p) \vel)
        + (\lag - m_1\partial_{m_1}\lag_1- m_2\partial_{m_2}\lag_2- m_1^d\partial_{m_1^d}\lag_1^d - p)\bnabla\bcdot\vel\\
        &\qquad + \partial_{\alpha_1}\lag_1 \: D_t \alpha_1+ \partial_{\alpha_2}\lag_2 \: D_t \alpha_2
        + \partial_{\rho_1^d}\lag_1 D_t \rho_1^d\\
        &\qquad + \partial_{m_1} \lag_1 (\partial_t m_1+\bnabla\bcdot(m_1\vel))
        + \partial_{m_2} \lag_2 (\partial_t m_2+\bnabla\bcdot(m_2\vel))\\
        &\qquad + \partial_{m_1^d} \lag_1^d (\partial_t m_1^d+\bnabla\bcdot(m_1^d\vel)).
    \end{aligned}
\end{equation}
The four last terms nullify thanks to the constraints while we develop $D_t \alpha_2$ using the volume occupation constraint $\alpha_1 + \alpha_2 + \alpha_1^d = 1$ and the conservation of $\alpha_1^d$.
It yields
\begin{equation}
    \label{eq:H_compute_3}
    \begin{aligned}
        0=\:&\partial_t \tEntropy +\bnabla\bcdot((\tEntropy+p) \vel) - (p+\lag^*)\bnabla\bcdot\vel
         + (\partial_{\alpha_1}\lag_1-\partial_{\alpha_2}\lag_2) \: D_t \alpha_1-\partial_{\alpha_2}\lag_2\: D_t \alpha_1^d\\[5pt]
        =\:&\partial_t \tEntropy +\bnabla\bcdot((\tEntropy+p) \vel) - (p+\lag^*-\alpha_1^d\partial_{\alpha_2}\lag_2)\bnabla\bcdot\vel+ (\partial_{\alpha_1}\lag_1-\partial_{\alpha_2}\lag_2) \: D_t \alpha_1.
    \end{aligned}
\end{equation}
Finally, the pressure equilibrium given by the second equation of \eqref{eq:null_variations} gives 
\begin{equation}
    \lag^*+\alpha_1^d\partial_{\alpha_2}\lag_2=-p,
\end{equation}
and we obtain the desired conservation equation on $\tEntropy$.

%% file: geomom_reconstruction.tex
\section{Large-scale and small-scale geometry description using GeoMOM geometric quantities}
\label{app:geomom_recon}

We propose here to illustrate why the choice of small-scale variables, which are at the same time moments of a kinetic modelling level as well as surface averaged geometric quantities is fundamentally convenient for two-scale modelling.
As their definitions are based on the surface-average operator in Section \ref{sec:polydisperse}, they can be computed for any flow topology
beyond any assumption on the geometry and dynamics of the interface.

We illustrate this point by comparing the evaluation of such quantities in a single realization of the collision of two droplets at small-scale. 

The simulation has been performed with the \texttt{ARCHER} code \citep{menard_coupling_2007} (see details in \cite{essadki_statistical_2019}) with $256\times256\times512$ cells in a volume $V_{tot}$ of $1\text{mm}\times1\text{mm}\times2\text{mm}$ and the geometric post-treatment of the obtained level-set is done with the open-source library \texttt{Mercur(v)e}, which relies on a triangulation of the interface as well as geometric properties preserving the topological invariants  \citep{cordesse_diffuse_2020}.
\begin{figure}
    \centering
    \begin{subfigure}[t]{.3\textwidth}
        \includegraphics[width=\textwidth]{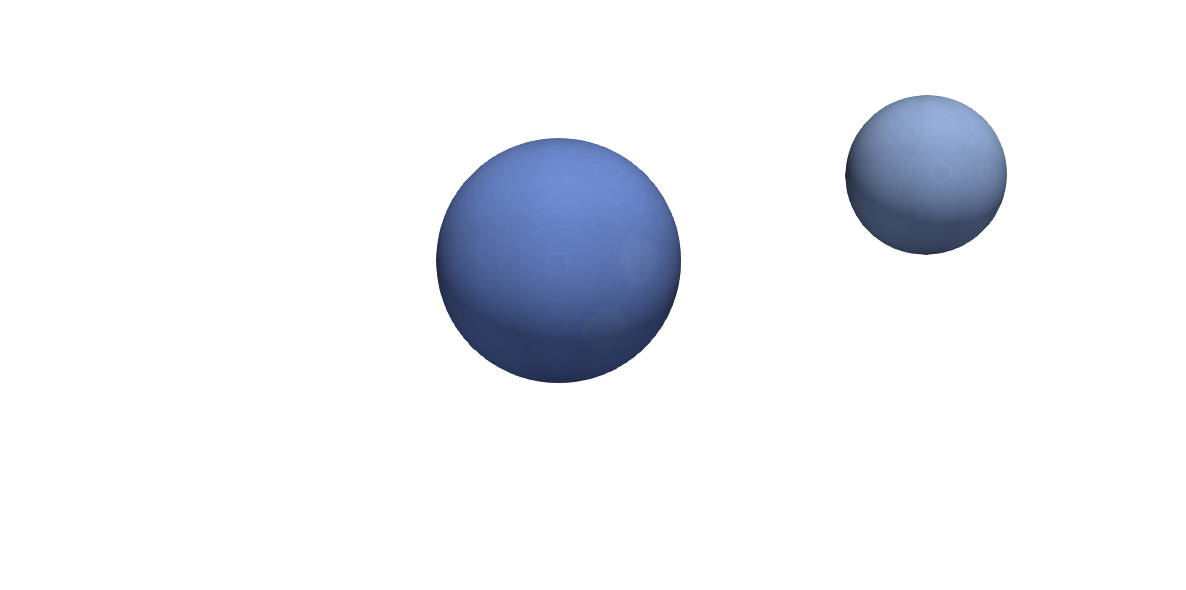}\\
        \caption{Before the collision.}
        \label{fig:before_coll}
    \end{subfigure}
    \begin{subfigure}[t]{.3\textwidth}
        \includegraphics[width=\textwidth]{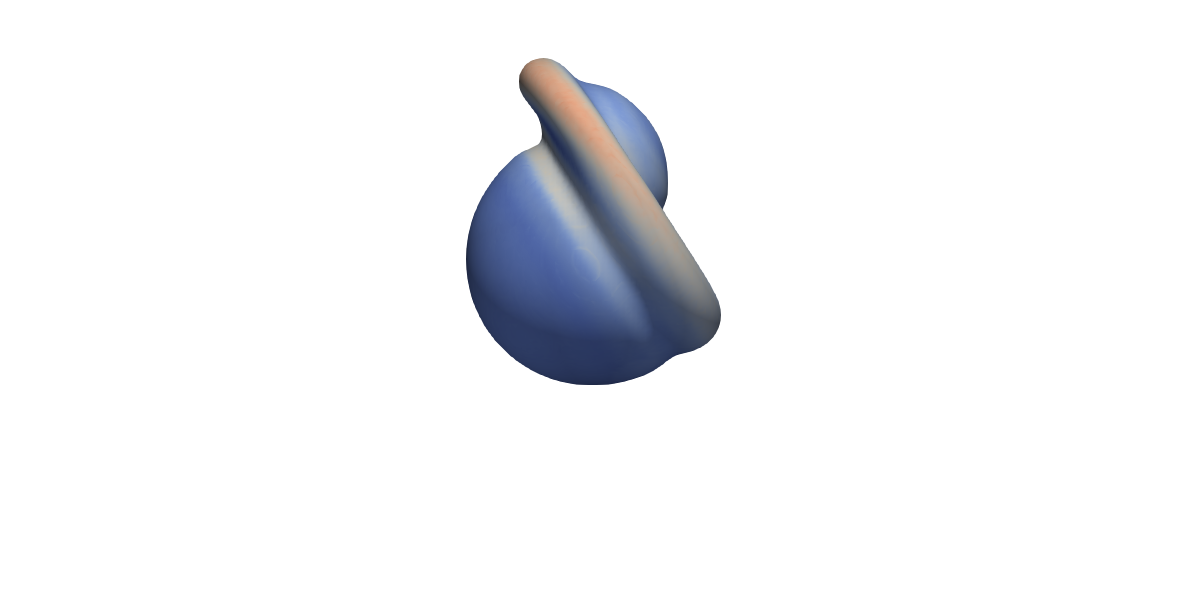}\\
        \caption{Beginning of the collision.}
        \label{fig:begin}
    \end{subfigure}
    \begin{subfigure}[t]{.3\textwidth}
        \includegraphics[width=\textwidth]{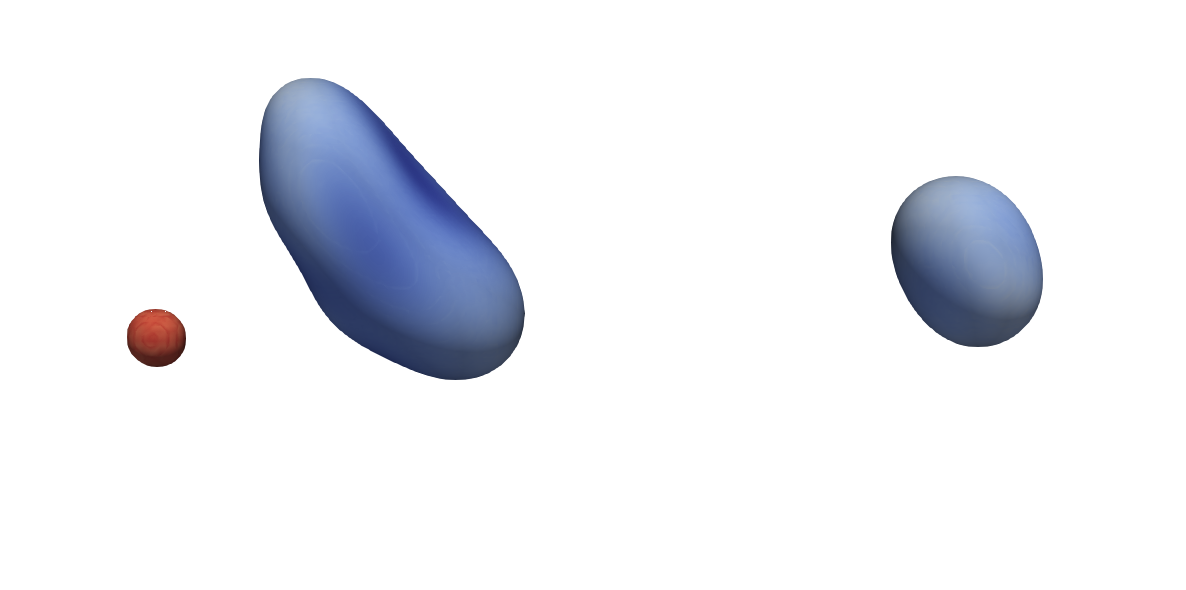}\\
        \caption{After the collision.}
        \label{fig:after_coll}
    \end{subfigure}
    \caption{Local mean curvature for the collision of two droplets that scales between $H_{min}=6.3 \cdot10^2\:\text{m}^{-1}$ (blue) and $H_{max}=2.4\cdot10^4\:\text{m}^{-1}$ (red).}
\end{figure}
Before the collision, the two droplets are spherical with diameters of $400\mu$m and $260\mu$m (see figure \ref{fig:before_coll}).
After the collision, there are three deformed inclusions with different sizes (see figure \ref{fig:after_coll}).

We usually consider a kinetic description at mesoscopic level based on a statistical ensemble average of several realizations, but for the purpose of the present illustration we stick to a single realization, since the kinetic description \eqref{eq:PBE_m} is also valid in this case of measure-valued number distribution functions.
At the kinetic level of description, we reconstruct the distribution using the two-point quadrature\footnote{We could also have used another reconstruction techniques such as the entropy maximization technique \cite{mead_maximum_1984,levermore_moment_1996,essadki_high_2018}, which maximizes the concave functional $n\mapsto -n \log n$, while constraining the values of some selected moments. Such techniques are better-suited for statistical descriptions corresponding to many realizations / inclusions or when a smooth NDF is required, such as when evaporation is taking place.}
\begin{equation}
    \label{eq:quadrature}
    n_{S_0}(\widehat{S}_0) = n_1 \delta(\widehat{S}_0-(S_0)_1) + n_2 \delta(\widehat{S}_0-(S_0)_2).
\end{equation}
We gather in tables \ref{tab:collision_geom} and \ref{tab:collision_weigths} the values of each geometric variable before and after collision, respectively from the analytical initial configuration and the geometric post-treatment, along with the corresponding weights and abscissas of \eqref{eq:quadrature}.
Let us study the reconstruction of the number density function is well-predicted through the chosen moments, showing how to interpret the chosen geometric variables.
Before the collision, \eqref{eq:quadrature} has the exact number of parameters to represent well the two droplets.
After the collision, droplets' sizes are sensitively changed with a droplet much smaller than the other two.
With \texttt{Mercur(v)e}, the geometric invariant associated surface-averaged Gauss curvature is well-preserved and recovers well the total number of droplets.
However, \eqref{eq:quadrature} has not enough parameters to detect three different sizes, even less the deformations.
However, deformations could be better modelled if we switched to  the eight-moment model. Nevertheless, even if it is a reduced-order model, the retained key information about the geometry at small-scale allow us to always have a representation, through quadrature, of a small-scale interface.

Let us conclude this illustration by underlining that the above results correspond to a treatment of the interface geometry as small-scale. We could also have taken the point of view of 
the large-scale, where all the geometry is not obtained through a geometric post-treatment, but can be locally estimated relying on $\overline{\alpha}_1$ only.
For instance, the interface area density and the mean curvature can be estimated using $\Vert\bnabla\overline{\alpha}_1\Vert$ and $ \bnabla\bcdot(\bnabla\overline{\alpha}_1/\Vert\bnabla\overline{\alpha}_1\Vert)$ \citep{goldman_curvature_2005}, as long as the characteristic lengths stay at large-scale, that is above a given threshold. 
\begin{table}
    \begin{subtable}{\textwidth}
        \centering
        \begin{tabular}{ccccc}
            & $\alpha_1^d$ & $\Sigma$ & $\SigH$ & $\SigG$ \\\hline
            Before coll. & $2.14 \cdot 10^{-2}$ & $3.57 \cdot 10^2\:\text{m}^{-1}$ & $2.1 \cdot 10^6\:\text{m}^{-2}$ & $1.3\cdot 10^{10}\:\text{m}^{-3}$\\
            After  coll. & $2.14 \cdot 10^{-2}$ & $3.91 \cdot 10^2\:\text{m}^{-1}$ & $2.5 \cdot 10^6\:\text{m}^{-2}$ & $1.9\cdot 10^{10}\:\text{m}^{-3}$
        \end{tabular}
        \caption{Geometric quantities before and after the collision.}
        \label{tab:collision_geom}
    \end{subtable}
    \begin{subtable}{\textwidth}
        \centering
        \vspace{.5cm}
        \begin{tabular}{ccccc}
            & $n_1\times V_{tot}$ & $(S_0)_1$ & $n_2\times V_{tot}$ & $(S_0)_2$\\\hline
            Before coll. & $1$ & $2.12 \cdot 10^{-7}\:\text{m}^2$ & $1$ & $5.03\cdot 10^{-7}\:\text{m}^2$ \\
            After  coll. & $0.76$ & $1.38\cdot 10^{-8}\:\text{m}^2$ & $2.24$ & $3.4\cdot 10^{-7}\:\text{m}^2$
        \end{tabular}
        \caption{Weights (nb. of droplets) in the domain of volume $V_{tot}$, and abscissas before and after the collision.}
        \label{tab:collision_weigths}
    \end{subtable}
\end{table}

%% file: SAP_poly.tex
\section{Derivation of the model for the polydisperse spray of incompressible droplets}
\label{app:SAP_polydisperse}

We consider here the two-scale model with capillarity at the small scale modelled by a polydisperse spray of spherical and incompressible droplets. The Lagrangian of the mixture is given by \eqref{eq:lagrangian_two_scale} with the modified small-scale Lagrangian given by \eqref{eq:lag_small_scale_capillarity} and reads
\begin{equation}
    \lag = \lag_1(\alpha_1, m_1, \vel) + \lag_2(\alpha_2, m_2, \vel) + \lag_1^d(m_1^d, \rho_1^d, z, \vel),
\end{equation}
where $z = \Sigma/m_1^d$ has a similar role to the variable defined in appendix \ref{app:incomp} as $D_t z = 0$. Moreover, $\rho_1^d$ is constrained by $D_t \rho_1^d=0$ and $\alpha_1^d$ is conserved such that the action, as defined in appendix \ref{app:SAP}, is decomposed according to each dependency,
\begin{align}
    \delta\Action_{\alpha_1} & = \int_\Omega \partial_{\alpha_1}\lag_1 \: \delta\alpha_1,\\
    \delta\Action_{\alpha_2} & = \int_\Omega \partial_{\alpha_2}\lag_2 \: \delta\alpha_2
    =-\int_{\Omega}\alpha_1^d\bnabla  (\partial_{\alpha_2}\lag_2) \bcdot \VarTraj 
    -\int_{\Omega}\partial_{\alpha_2}\lag_2 \: \delta \alpha_1,\\
    \delta\Action_{m_k} & = \int_\Omega \partial_{m_k}\lag_k  \: \delta m_k = \int_\Omega m_k \bnabla (\partial_{m_k}\lag_k)\bcdot \VarTraj,\\
    \delta\Action_{\rho_1^d} & = \int_\Omega -\partial_{\rho_1^d}\lag_1^d \: \bnabla \rho_1^d \bcdot \VarTraj,\\
    \delta\Action_{z} & = \int_\Omega -\partial_{z}\lag_1^d \: \bnabla z \bcdot \VarTraj,\\
    \delta\Action_{\vel} & = \int_\Omega -\left(\partial_t\boldsymbol{K}+\bnabla\bcdot(\boldsymbol{K} \otimes \vel)+\boldsymbol{K}\bcdot\bnabla\vel\right)\bcdot\VarTraj,
\end{align}
The variation of the total action then writes
\begin{equation}\label{eq:action_variation_with_capillarity}
    \begin{aligned}
        \delta \Action =&-\int_{\Omega}\left(\partial_t\boldsymbol{K}+\bnabla\bcdot(\boldsymbol{K} \otimes \vel)+\boldsymbol{K}\bcdot\bnabla\vel+\alpha_1^d\bnabla  (\partial_{\alpha_2}\lag_2)+\partial_{z}\lag_1^d\:\bnabla z \right.\\
        &\qquad\left.-m_1 \bnabla (\partial_{m_1}\lag_1)-m_2 \bnabla (\partial_{m_2}\lag_2)-m_1^d \bnabla (\partial_{m_1^d}\lag_1^d)\right)\bcdot\VarTraj\\
        &+\int_{\Omega}(\partial_{\alpha_1}\lag_1-\partial_{\alpha_2}\lag_2) \: \delta \alpha_1,\\
        =& -\int_{\Omega} \left(\partial_t \boldsymbol{K}+\bnabla\bcdot(\boldsymbol{K} \otimes \vel) - \bnabla(\lag^*-\alpha_1^d\partial_{\alpha_2}\lag_2)
        -(\partial_{\alpha_1}\lag_1-\partial_{\alpha_2}\lag_2)\bnabla\alpha_1\right)\bcdot \VarTraj\\
        & +\int_{\Omega}(\partial_{\alpha_1}\lag_1-\partial_{\alpha_2}\lag_2)\:\delta\alpha_1.
    \end{aligned}
\end{equation}
Nullifying the variations gives the following system
\begin{equation}
    \label{eq:SAP_eq_small_scale_polydisperse}
    \begin{cases}
        \partial_t\boldsymbol{K}+\bnabla\bcdot(\boldsymbol{K} \otimes \vel) - \bnabla(\lag^*-\alpha_1^d\partial_{\alpha_2}\lag_2) = 0, \\
        \partial_{\alpha_1}\lag_1-\partial_{\alpha_2}\lag_2 = 0.
    \end{cases}
\end{equation}
We replace then the derivatives of the Lagrangian by their expressions
\begin{equation}
    \begin{gathered}
        \partial_{\vel}\lag=\boldsymbol{K}=\rho \vel,
        \qquad
        \lag^* =-\alpha_1 p_1-\alpha_2p_2,
        \qquad
        \partial_{\alpha_2}\lag_2 =p_2,
        \qquad
        \partial_{\alpha_1}\lag_1 = p_1.
    \end{gathered}
\end{equation}
Then, \eqref{eq:SAP_eq_small_scale_polydisperse} becomes
\begin{equation}
    \begin{cases}
        \partial_t(\rho\vel)+\bnabla\bcdot(\rho\vel \otimes \vel + p \boldsymbol{I}) = 0, \\
        p:=p_1=p_2.
    \end{cases}
\end{equation}

%% file: diff_geom_S_SH.tex
\section{Results of differential geometry}
\label{app:diff_geom}

We recall here some elements to derive the evolution in time of the geometry for closed inclusions using the tools of differential geometry.
The reader is referred to \cite{kreyszig_differential_1991} for a general introduction to differential geometry and to \cite{capovilla_deformations_2003,deserno_fluid_2015} for a briefer introduction and the derivation of the formulas presented hereafter.

\subsection{First- and second-order variations of a closed inclusion}

Let us consider the deformation of a closed inclusion defined by its position vector mapped by $\mathcal{U}$
\begin{equation}
    \boldsymbol{r}(u,v) = \boldsymbol{r}_0(u,v) + \delta \boldsymbol{r}(u,v),
    \qquad
    (u,v)\in\mathcal{U},
\end{equation}
where $\boldsymbol{r}_0$ is the position before the deformation and $\delta \boldsymbol{r}$ the deformation vector that we assume to be small.
From now on, we drop the dependencies on $(u,v)$ to lighten the formulas.
For convex and closed inclusions, we write the deformation in the direction of the non-deformed surface normal,
\begin{equation}
    \boldsymbol{r} = \boldsymbol{r}_0 + \psi \boldsymbol{n},
\end{equation}
where $\boldsymbol{n}:=(\boldsymbol{e}_{\theta}\times\boldsymbol{e}_{\phi})/\Vert\boldsymbol{e}_{\theta}\times\boldsymbol{e}_{\phi}\Vert$ is the local normal and $\boldsymbol{e}_a:=\partial_a\boldsymbol{r}$ the tangential vectors and $\psi$ the local amplitude of the deformation.
Then, we are interested in the first-order and second-order variations of particular geometric quantities of the whole inclusion.
With the surface-weighted average operator $\widetilde{(\cdot)}$ defined in \eqref{eq:average_op} for one inclusion, we focus on the volume $V$, the surface area $S$ and the surface-weighted mean curvature $S\tilde{H}$.
There is no variation associated with the surface-weighted Gauss curvature $S\tilde{G}$ as a result of the Gauss-Bonnet theorem \citep{kreyszig_differential_1991}.
The variations of these quantities up to second order in $\psi$ are found in \cite{capovilla_deformations_2003} and written with Einstein summation rule for implicit indexes $a,b \in\{u,v\}$,
\begin{equation}
    \label{eq:geom_diff_quantities}
    \begin{aligned}
        \delta V &= \int_{\mathcal{U}}
        (
            \psi
            +H_0\psi^2
        )
        A_0 + o(\psi^2),\\
        \delta S &= \int_{\mathcal{U}}
        (
            2H_0\psi
            +G_0\psi^2
            -\tfrac{1}{2}\psi\Delta\psi
        )
        A_0 + o(\psi^2),\\
        \delta (S\tilde{H}) &= \int_{\mathcal{U}}
        (
            G_0\psi
            +\tfrac{1}{2}\psi[(\boldsymbol{K}^{ab}-2H_0\boldsymbol{g}^{ab}):(\bnabla_{a}\otimes\bnabla_{b})\psi-\boldsymbol{G}:\boldsymbol{K}^{ab}\psi]
        )
        A_0 + o(\psi^2),
    \end{aligned}
\end{equation}
where $H_0$ and $G_0$ are the non-deformed mean and Gauss curvature, $\Delta$ is the surface Laplacian, $\boldsymbol{K}^{ab}$ the extrinsic curvature tensor, $\boldsymbol{g}^{ab}$ the inverse of the metric tensor, $\bnabla_a$ is the vector of covariant derivative with respect to $a$, and $\boldsymbol{G}$ the Einstein tensor.

\subsection{Incompressible oscillation of the sphere}

In Section \ref{sec:kinetic-model-oscill}, we are specifically interested in a sphere deformed by the axisymmetric second spherical harmonic.
We choose the spherical coordinates $(u,v)=(\theta,\phi)$ on $\mathcal{U} = (0,\pi)\times(0,2\pi)$, and we denote $\boldsymbol{e}_r=(\sin\theta\cos\phi, \sin\theta\sin\phi,\cos\theta)$ such that
\begin{equation}
    \boldsymbol{r}_0 = R_0 \boldsymbol{e}_r,
    \qquad
    \psi(\theta,\phi) = x_0 Y_0 + x_2 Y_2(\theta),
\end{equation}
where $R_0$ is the constant radius of the non-deformed sphere, $Y_0$ is the zeroth spherical harmonic that corresponds to the isotropic perturbation that ensures incompressibility, and $Y_2$ is the second spherical harmonic associated with the oscillatory motion.
Then, we evaluate the geometric quantities in \eqref{eq:geom_diff_quantities},
\begin{equation}
    \label{eq:spherical_eval_geom_diff}
    \begin{gathered}
        \boldsymbol{n} = \boldsymbol{e}_r,
        \quad
        H_0 = R_0^{-1},
        \quad
        G_0 = R_0^{-2},
        \quad
        A_0 = R_0^2A_{\mathbb{S}^2},\\
        \quad
        \boldsymbol{K}^{ab} = R_0^{-1}\boldsymbol{g}^{ab},
        \quad
        \boldsymbol{g}^{ab}:\bnabla_{a}\otimes\bnabla_{b} = \Delta,
        \quad
        \Delta = R_0^{-2}\Delta_{\mathbb{S}^2},
        \quad
        \boldsymbol{G} = \boldsymbol{0},
    \end{gathered}
\end{equation}
where $\Delta_{\mathbb{S}^2}$ and $A_{\mathbb{S}^2}$ are the surface Laplacian and the surface element on the unit sphere $\mathbb{S}^2$.
Furthermore, the orthonormal spherical harmonics $Y_0$ and $Y_2$ are eigenfunctions of the spherical Laplacian $\Delta_{\mathbb{S}^2}$ such that
\begin{equation}
    \label{eq:spherical_harm_prop}
    \Delta_{\mathbb{S}^2} Y_k = -k(k+1) Y_k,
    \qquad
    \int_{\mathcal{U}} Y_kY_{k^{\prime}} A_{\mathbb{S}^2}= \delta_{k,k^{\prime}}.
\end{equation}
Evaluating the variations of $V$, $S$ and $S\tilde{H}$ up to second order in $x_0$ and $x_2$ yields
\begin{equation}
    \begin{aligned}
        \delta V &= R_0^2 x_0 + R_0 x_0^2 + R_0x_2^2 + o(x_0^2+x_2^2),\\
        \delta S &= 2 R_0 x_0 + x_0^2 + 4x_2^2 + o(x_0^2+x_2^2),\\
        \delta (S\tilde{H}) &= x_0 + 3R_0^{-1}x_2^2 + o(x_0^2+x_2^2).
    \end{aligned}
\end{equation}
As we consider incompressible oscillations $\delta V=0$, this links the dynamics of $x_0$ to the one of $x_2$
\begin{equation}
    \delta V = 0
    \quad
    \Rightarrow
    \quad
    x_0 = -R_0^{-1}x_2^2 + o(x_2^2).
\end{equation}
The variations of the geometric quantities then simplify to
\begin{equation}
    \label{eq:final_variations}
    \delta V = 0,
    \quad
    \delta S = 2x_2^2 + o(x_2^2),
    \quad
    \delta (S\tilde{H}) = 2R_0^{-1}x_2^2 + o(x_2^2).
\end{equation}
We then obtain equations \eqref{eq:def_oscillation_quantities} by introducing $S_0 = 4\pi R_0^2$ and $\chi = x_2(2/S_0)^{1/2}$.

\subsection{Variations of the oriented geometric quantities}

In Section \ref{sec:org_osc}, new geometric quantities $S_{\parallel}$, $S_{\perp}$, $S_{\parallel}\widetilde{H}^{\parallel}$ and $S_{\perp}\widetilde{H}^{\perp}$ are defined to recover first order variations in $x_2$ using the decomposition of the local surface element into
\begin{equation}
    A = \Vert \boldsymbol{e}_{\theta} \times \boldsymbol{e}_{\phi} \Vert
    =
    \left| (\boldsymbol{e}_{\theta} \times \boldsymbol{e}_{\phi}) \bcdot \boldsymbol{N} \right| 
    +\left(\Vert \boldsymbol{e}_{\theta} \times \boldsymbol{e}_{\phi} \Vert - \left| (\boldsymbol{e}_{\theta} \times \boldsymbol{e}_{\phi}) \bcdot \boldsymbol{N} \right|\right)
    =:A_{\parallel} + A_{\perp},
\end{equation}
where $\boldsymbol{e}_a:=\partial_a \boldsymbol{r}$ for $a=\theta,\phi$, and $\boldsymbol{e}_v:=\partial_v \boldsymbol{r}(u,v)$ are the tangential vectors and $\boldsymbol{N}=(0,0,1)$ is a constant vector chosen along the axisymmetric axis.
Then for any local geometric quantity $X(u,v)$, we split the variation of the surface-averaged geometric quantity into 
\begin{equation}
    \delta(S\widetilde{X}) = \delta(S_{\parallel}\widetilde{X}^{\parallel})
    +\delta(S_{\perp}\widetilde{X}^{\perp}),
\end{equation}
such that we can focus on the variation $\delta(S_{\parallel}\widetilde{X}^{\parallel})$, and $\delta(S_{\perp}\widetilde{X}^{\perp})$ follows from \eqref{eq:final_variations}. The variation of $S_{\parallel}\widetilde{X}^{\parallel}$ reads
\begin{equation}
    \delta(S_{\parallel}\widetilde{X}^{\parallel}) = \int_{\mathcal{U}}\delta(XA_{\parallel})
    = \int_{\mathcal{U}}\delta\left(X\left| (\boldsymbol{e}_{\theta} \times \boldsymbol{e}_{\phi}) \bcdot \boldsymbol{N} \right|\right),
\end{equation}
We get rid of the absolute value by remarking that the perturbations along harmonics $Y_0$ and $Y_2$ are symmetric with respect to the equatorial plane of the droplet.
We split $\mathcal{U}$ into two hemispheres using the half unit sphere mapping $\tfrac{1}{2}\mathcal{U} = (0,\pi/2)\times(0,2\pi)$ where $(\boldsymbol{e}_{\theta} \times \boldsymbol{e}_{\phi} )\bcdot \boldsymbol{N}>0$ and perform a change of variables leading to
\begin{equation}
    \begin{aligned}
        \delta(S_{\parallel}\widetilde{X}^{\parallel})
        &= \int_{\tfrac{1}{2}\mathcal{U}}\delta(X (\boldsymbol{e}_{\theta} \times \boldsymbol{e}_{\phi}) \bcdot \boldsymbol{N} )
        - \int_{\mathcal{U}\setminus\tfrac{1}{2}\mathcal{U}}\delta(X (\boldsymbol{e}_{\theta} \times \boldsymbol{e}_{\phi}) \bcdot \boldsymbol{N})\\
        &= 2\int_{\tfrac{1}{2}\mathcal{U}}\delta(X(\boldsymbol{e}_{\theta} \times \boldsymbol{e}_{\phi})) \bcdot \boldsymbol{N}.
    \end{aligned}
\end{equation}
Then, the variation is decomposed following
\begin{equation}
    \label{eq:oriented_quantity}
    \delta(S_{\parallel}\widetilde{X}^{\parallel}) 
    = 2\int_{\tfrac{1}{2}\mathcal{U}}\delta(XA)(\boldsymbol{n}\bcdot\boldsymbol{N}) + 
    2 \int_{\tfrac{1}{2}\mathcal{U}}X_0A_0\delta\boldsymbol{n}\cdot\boldsymbol{N},
\end{equation}
with $A_0\delta\boldsymbol{n} = \delta(\boldsymbol{e}_{\theta} \times \boldsymbol{e}_{\phi}) - \boldsymbol{n}\delta A$.
The first-order variation in $x_2$ is non-trivial here, and we retain only the first-order terms in $\psi$ for $\delta(HA)$ and $\delta A$ in \eqref{eq:geom_diff_quantities}. Only the first-order variation of $\delta(\boldsymbol{e}_{\theta} \times \boldsymbol{e}_{\phi})$ is still undetermined. For the first-order deformation $\psi = x_2 Y_2 + o(x_2)$, it yields
\begin{equation}
    \begin{aligned}
        \delta(\boldsymbol{e}_{\theta} \times \boldsymbol{e}_{\phi})
        &= 
        \partial_{\theta}\boldsymbol{r} \times \partial_{\phi}\boldsymbol{r} - \partial_{\theta}\boldsymbol{r}_0 \times \partial_{\phi}\boldsymbol{r}_0\\
        &= 
        x_2R_0Y_2
        (\partial_{\theta}\boldsymbol{n}\times \partial_{\phi}\boldsymbol{n})+x_2R_0(\partial_{\theta}Y_2)(\boldsymbol{n}\times \partial_{\phi}\boldsymbol{n})
        +R_0x_2Y_2(\partial_{\theta}\boldsymbol{n} \times \partial_{\phi}\boldsymbol{n})+ o(x_2^2).
    \end{aligned}
\end{equation}
For the sphere, we recall that 
\begin{equation}
    \begin{gathered}
        \partial_{\theta}\boldsymbol{n} = R_0^{-1} \boldsymbol{e}_{\theta},
        \quad
        \partial_{\phi}\boldsymbol{n} = R_0^{-1} \boldsymbol{e}_{\phi},
        \quad
        \boldsymbol{e}_{\theta}\times\boldsymbol{e}_{\phi} = A_0 \boldsymbol{n},\\
        A_0=R_0^2\sin\theta,
        \quad
        \boldsymbol{N}\bcdot\boldsymbol{n}=\cos\theta,
        \quad
        \boldsymbol{N}\bcdot\boldsymbol{e}_{\theta}=-R_0\sin\theta.
    \end{gathered}
\end{equation}
The first-order variation $\delta(\boldsymbol{e}_{\theta} \times \boldsymbol{e}_{\phi})$ then reads
\begin{equation}
    \begin{aligned}
        \delta(\boldsymbol{e}_{\theta} \times \boldsymbol{e}_{\phi})
        &= 
        x_2R_0^{-1}Y_2
        (\boldsymbol{e}_{\theta}\times \boldsymbol{e}_{\phi})
        +x_2(\partial_{\theta}Y_2)(\boldsymbol{n}\times \boldsymbol{e}_{\phi})
        +R_0^{-1}x_2Y_2(\boldsymbol{e}_{\theta} \times \boldsymbol{e}_{\phi})+ o(x_2^2)\\
        &= 
        x_2R_0^{-1}Y_2A_0\boldsymbol{n}-x_2(\partial_{\theta}Y_2)A_0R_0^{-2}\boldsymbol{e}_{\theta} +R_0^{-1}x_2Y_2A_0\boldsymbol{n}+ o(x_2^2)\\
        &=2x_2R_0^{-1}Y_2A_0\boldsymbol{n}-x_2(\partial_{\theta}Y_2)\boldsymbol{e}_{\theta}+ o(x_2^2).
    \end{aligned}
\end{equation}
Now, taking $X=1$ in \eqref{eq:oriented_quantity} gives
\begin{equation}
    \begin{aligned}
        \delta S_{\parallel}
        &= 2 \int_{\tfrac{1}{2}\mathcal{U}}
        \delta(\boldsymbol{e}_u \times \boldsymbol{e}_v)\bcdot\boldsymbol{N}\\
        &= 2 \int_{\tfrac{1}{2}\mathcal{U}}
        2x_2R_0^{-1}Y_2A_0(\boldsymbol{n}\bcdot\boldsymbol{N})-x_2A_0R_0^{-2}(\partial_{\theta}Y_2)(\boldsymbol{e}_{\theta}\bcdot\boldsymbol{N})+ o(x_2^2)\\
        &= 4x_2R_0 \int_{\tfrac{1}{2}\mathcal{U}}
        Y_2\sin\theta\cos\theta
        +2x_2 R_0\int_{\tfrac{1}{2}\mathcal{U}}(\partial_{\theta}Y_2)\sin^2\theta+ o(x_2^2)\\
        &= 4x_2R_0 \frac{\sqrt{5 \pi }}{8}
        +2x_2 R_0\left(-\frac{3 \sqrt{5 \pi }}{4}\right)+ o(x_2^2)\\
        &= -x_2R_0 \sqrt{5\pi}+ o(x_2^2).
    \end{aligned}
\end{equation}
For $X=H$, there is an extra term in the first-order variation $\delta(HA)=(G_0 -\tfrac{1}{2}\Delta)\psi A_0$ \cite{capovilla_deformations_2003} which has been nullified in \eqref{eq:geom_diff_quantities} as the inclusion is closed.
Here, it is taken into account as we integrate twice over a half inclusion.
It yields
\begin{equation}
    \begin{aligned}
        \delta S_{\parallel}\tilde{H}^{\parallel}
        &= 2\int_{\tfrac{1}{2}\mathcal{U}}\delta(HA)(\boldsymbol{n}\bcdot\boldsymbol{N}) + 
        2 \int_{\tfrac{1}{2}\mathcal{U}}H_0(\delta(\boldsymbol{e}_{\theta} \times \boldsymbol{e}_{\phi}) - \boldsymbol{n}\delta A)\cdot\boldsymbol{N},\\
        &= 2\int_{\tfrac{1}{2}\mathcal{U}}(G_0-2H_0^2)\psi A_0(\boldsymbol{n}\bcdot\boldsymbol{N}) + 
        \int_{\tfrac{1}{2}\mathcal{U}}\Delta\psi A_0(\boldsymbol{n}\bcdot\boldsymbol{N}) +
        2 \int_{\tfrac{1}{2}\mathcal{U}}H_0\delta(\boldsymbol{e}_{\theta} \times \boldsymbol{e}_{\phi})\cdot\boldsymbol{N},\\
        &= -2x_2\int_{\tfrac{1}{2}\mathcal{U}}Y_2 \sin\theta\cos\theta + 
        -x_2 \int_{\tfrac{1}{2}\mathcal{U}}(\Delta_{\mathbb{S}^2} Y_2) \sin\theta\cos\theta
        +
        2 R_0^{-1}\int_{\tfrac{1}{2}\mathcal{U}}\delta(\boldsymbol{e}_{\theta} \times \boldsymbol{e}_{\phi})\cdot\boldsymbol{N},\\
        &= -x_2 \frac{1}{2}\sqrt{5 \pi }.
    \end{aligned}
\end{equation}
Replacing $x_2$ with its expression in $\chi$ and $S_0$ yields geometric relations \eqref{eq:def_geo_quant_oscil_paral_perp}.

%% file: poly_quad.tex
\section{Weights and quadrature points of the bi-disperse quadrature}
\label{app:polydisperse-closure}

The bi-disperse closure for $n_1, n_2,(S_0)_1, (S_0)_2$ in terms of the moment in size only $M_k=M_{k,0,0}^\xi$ with $k=0,1/2,1,3/2$ is obtained by solving the truncated moment problem with Mathematica \citep{wolfram_research_mathematica_2023} and reads
\begin{equation}
    \begin{aligned}
        n_i &= \frac{1}{2}\left(M_0+(-1)^{i+1}\frac{3M_0M_1M_{1/2}-2M_{1/2}^3-M_0^2M_{3/2}}{\sqrt{\Delta}}\right),\\
        (S_0)_i & = (2(M_{1/2}^2-M_0M_1)^2)^{-1}
        \left(M_0^2M_{3/2}^2-M_1^2M_{1/2}^2+2(M_0M_1^3+M_{1/2}^3M_{3/2})\right.\\
        &\qquad\left.-4M_0M_{1/2}M_1M_{3/2}+(-1)^{i+1}(M_0M_{3/2}-M_1M_{1/2})\sqrt{\Delta}\right),\\
        \text{with}
        \quad \Delta & = 4 M_0M_1^3-3M_1^2M_{1/2}^2-6M_0M_{1/2}M_1M_{3/2}+4M_{1/2}^3M_{3/2}+M_0^2M_{3/2}^2.
    \end{aligned}
\end{equation}
It can be written with the geometric variables using either relations \eqref{eq:geom_var_moments_rel_S}, \eqref{eq:geomom_osc_rel}-\eqref{eq:geomom_osc_rel_decomp} or \eqref{eq:rel_mom_geom_osc_sync}-\eqref{eq:rel_mom_geom_osc_sync_decomp}.
Mathematica also shows that these relations yield positive values of $n_i$ and $(S_0)_i$ provided that the moments $M_k$ are positive and
\begin{equation}
    M_{1/2}M_{3/2}-M_1^2>0,
    \qquad
    M_0M_1-M_{1/2}^2>0.
\end{equation}
These last two conditions ensure the positivity of Hankel matrices involved in the realizability conditions of the Hausdorff truncated moment problem \citep{schmudgen_moment_2017}.

%% file: SAP_poly_osc.tex
\section{Hamilton's SAP for the polydisperse spray of oscillating droplets}
\label{app:SAP_osc}

This model is built on the basis of the two-scale model of Section \ref{sec:two-scale} where additional energies are added to take into account capillarity at the small-scale along with the internal flow of the droplets. We recall the Lagrangian given in \eqref{eq:lagrangian_oscill} for the two-scale mixture
\begin{equation}
    \begin{gathered}
        \lag=
        \lag_1\left(\alpha_1, m_1, \vel\right)
        +\lag_2\left(\alpha_2, m_2, \vel\right)
        +\lag_{1}^d\left(m_1^d, \rho_1^d, \vel\right)\\
        +\lag^{vib}_1\left(n_1, (S_0)_1, \chi_1, \dot{\chi}_1\right)
        +\lag^{vib}_2\left(n_2, (S_0)_2, \chi_2, \dot{\chi}_1\right),
    \end{gathered}
\end{equation}
where $\lag^{vib}_i =\frac{1}{2}\nu n_i (S_0)_i^{5/2}\chi_i^2 - \sigma n_i (S_0)_i - \frac{1}{2}\gamma n_i (S_0)_i \chi_i^2$. We define the action $\mathcal{A}$ associated to the Lagrangian similarly to the one of appendix \ref{app:SAP}. As same as previous models, $\alpha_1$ is a free variable in the minimization process, while effective densities $m_k$ are conserved, $\rho_1^d$ is advected. For the additional variables, the number densities of droplets $n_i$ are conserved, the surfaces $(S_0)_i$ are advected, $\chi_i$ are free variables describing the oscillatory motion of the droplets, and $\dot{\chi}_i$ are linked to time derivatives of $\chi_i$ with $D_t \chi_i = \dot{\chi}_i$. This last constraint translates in terms of variations
\begin{equation}
    \delta(\dot{\chi}_i)=\delta(D_t\chi_i)=\partial_t (\delta\chi_i)+\vel \bcdot \bnabla (\delta \chi_i) + \delta \vel \bcdot \bnabla \chi_i.
\end{equation}
Denoting $K_{\dot{\chi}_i}=\partial_{\dot{\chi}_i}\lag^{vib}_i$ and $\boldsymbol{\vel} = \partial_{\vel}\lag$, we decompose then the variation of the action according to each dependency
\begin{align}
    \delta\Action_{\alpha_1} & = \int_\Omega \partial_{\alpha_1}\lag_1 \: \delta\alpha_1,\\
    \delta\Action_{\alpha_2} & = \int_\Omega \partial_{\alpha_2}\lag_2 \: \delta\alpha_2
    =-\int_{\Omega}\alpha_1^d\bnabla  (\partial_{\alpha_2}\lag_2) \bcdot \VarTraj 
    -\int_{\Omega}\partial_{\alpha_2}\lag_2 \: \delta \alpha_1,\\
    \delta\Action_{m_k} & = \int_\Omega \partial_{m_k}\lag_k  \: \delta m_k = \int_\Omega m_k \bnabla (\partial_{m_k}\lag_k)\bcdot \VarTraj,\\
    \delta\Action_{\rho_1^d} & = \int_\Omega -\partial_{\rho_1^d}(\lag_1^d+\lag^{vib}_1+\lag^{vib}_2) \: \bnabla \rho_1^d \bcdot \VarTraj,\\
    \delta\Action_{(S_0)_i} & = \int_\Omega -\partial_{(S_0)_i}\lag^{vib}_i\: \bnabla (S_0)_i \bcdot \VarTraj,\\
    \delta\Action_{n_i} & = \int_\Omega \partial_{n_i}\lag^{vib}_i  \: \delta n_i = \int_\Omega n_i \bnabla (\partial_{n_i}\lag^{vib}_i)\bcdot \VarTraj,\\
    \delta\Action_{\chi_i} & = \int_\Omega \partial_{\chi_i}\lag^{vib}_i  \: \delta \chi_i,\\
    \delta\Action_{\dot{\chi}_i} & = \int_\Omega K_{\dot{\chi}_i}  \: \delta \dot{\chi}_i=\int_\Omega K_{\dot{\chi}_i} (\partial_t (\delta\chi_i)+\vel \bcdot \bnabla (\delta \chi_i) + \delta \vel \bcdot \bnabla \chi_i),\\
    & = 
    -\int_\Omega (\partial_t K_{\dot{\chi}_i} + \bnabla\bcdot(K_{\dot{\chi}_i}\vel))\delta\chi_i\\
    &-\int_\Omega   (\partial_t(K_{\dot{\chi}_i}\bnabla \chi_i)+\bnabla\bcdot ((K_{\dot{\chi}_i}\bnabla \chi_i)\vel)+K_{\dot{\chi}_i}\bnabla \chi_i\bcdot\bnabla\vel)\bcdot\boldsymbol{\VarTraj},\\
    \delta\Action_{\vel} & = \int_\Omega -\left(\partial_t\boldsymbol{K}+\bnabla\bcdot(\boldsymbol{K} \otimes \vel)+\boldsymbol{K}\bcdot\bnabla\vel\right)\bcdot\VarTraj.
\end{align}
We denote $\lag_k^* = m_k\partial_{m_k}\lag_k-\lag_k$, $\lag_i^{vib,*} = n_i\partial_{n_i}\lag_i^{vib}-\lag_i^{vib}$ and $\lag^*=\lag_1^*+\lag_2^*+\lag_1^{d,*}$ such that the variation of the action related to the mixture lagrangian reads
\begin{equation}
    \delta \Action = \int 
    \boldsymbol{A}_{\VarTraj} \bcdot \VarTraj 
    +\boldsymbol{A}_{\alpha_1} \delta \alpha_1 
    +\boldsymbol{A}_{\chi_1} \delta\chi_1
    +\boldsymbol{A}_{\chi_2} \delta\chi_2,
\end{equation}
with
\begin{equation}
    \begin{aligned}
        \boldsymbol{A}_{\alpha_1} & = 
        \partial_{\alpha_1}\lag_1
        -\partial_{\alpha_2}\lag_2,\\
        \boldsymbol{A}_{\chi_i} & = \partial_{\chi_i}\lag_{k}^{vib} - \partial_t K_{\dot{\chi}_i} - \bnabla \bcdot (K_{\dot{\chi}_i} \vel),\\
        \boldsymbol{A}_{\VarTraj} & =
        -\left(
            \partial_t \boldsymbol{K} + \bnabla \bcdot (\boldsymbol{K} \otimes \vel) 
            -\bnabla (\lag^*-\alpha_1^d\partial_{\alpha_2}\lag_2)
            -\boldsymbol{A}_{\alpha_1}\bnabla\alpha_1
            -\boldsymbol{A}_{\chi_1}\bnabla \chi_1
            -\boldsymbol{A}_{\chi_2}\bnabla \chi_2
        \right)
    \end{aligned}
\end{equation}
and $\lag^{vib,*}_i:=n_i\partial_{n_i}\lag^{vib}_i-\lag^{vib}_i$.
Nullifying the variations gives the following system
\begin{equation}
    \begin{cases}
        \partial_t \boldsymbol{K} + \bnabla \bcdot (\boldsymbol{K} \otimes \vel) 
        -\bnabla (\lag^*-\alpha_1^d\partial_{\alpha_2}\lag_2)=0,\\
        \partial_{\chi_1}\lag_{1}^{vib} - \partial_t K_{\dot{\chi}_1} - \bnabla \bcdot (K_{\dot{\chi}_1} \vel)=0,\\
        \partial_{\chi_2}\lag_{2}^{vib} - \partial_t K_{\dot{\chi}_2} - \bnabla \bcdot (K_{\dot{\chi}_2} \vel)=0,\\
        \partial_{\alpha_1}\lag_1
        -\partial_{\alpha_2}\lag_2=0.
    \end{cases}
\end{equation}
Evaluating the derivatives of the Lagrangian as defined in \eqref{eq:lagrangian_osc_small_scale} yields
\begin{equation}
    \begin{gathered}
        \boldsymbol{K}_{\vel}=\partial_{\vel}\lag=\rho \vel,
        \qquad
        \lag^* =-\alpha_1 p_1-\alpha_2p_2,
        \qquad
        \partial_{\alpha_2}\lag_2 =p_2,
        \qquad
        \partial_{\alpha_1}\lag_1 = p_1,\\
        \partial_{\chi_i}\lag_i^{vib} = -\gamma n_i (S_0)_i \chi_i,
        \qquad
        K_{\dot{\chi}_i} = \partial_{\dot{\chi}_i}\lag_i^{vib} = \nu n_i (S_0)_i^{5/2}\dot{\chi}_i,
    \end{gathered}
\end{equation}
Finally, with the constraints and the relation $D_t \chi_i = \dot{\chi}_i$, we write the final system in its conservative form
\begin{equation}
    \begin{cases}
        \setlength{\arraycolsep}{1pt}
        \begin{array}{llll}
            \partial_t m_k &+ \bnabla \bcdot (m_k \vel)&=0,   &\qquad k=1,2,1^d,\\
            \partial_t n_i &+ \bnabla \bcdot (n_i \vel)&=0,   &\qquad i=1,2,\\
            \partial_t (n_i(S_0)_i) &+ \bnabla \bcdot (n_i(S_0)_i \vel) &=0,   &\qquad i=1,2,\\
            \partial_t(n_i\dot{\chi}_i)&+ \bnabla \bcdot (n_i\dot{\chi}_i \vel)&=-\omega_i^2 n_i\chi_i, &\qquad i=1,2, \\
            \partial_t(n_i\chi_i)&+ \bnabla \bcdot (n_i\chi_i \vel) &=n_i\dot{\chi}_i, &\qquad i=1,2,\\
            \partial_t (\rho \vel) &+ \bnabla \bcdot (\rho \vel \otimes \vel+p \boldsymbol{I})&=0, &                    
        \end{array}
    \end{cases}
\end{equation}
where $p:=p_1=p_2$ and $\omega_i^2 = \gamma/\nu (S_0)_i^{-3/2}$.

%% file: monodisperse.tex
\section{The monodisperse case}
\label{app:monodisperse}

Even though the polydispersion is a key feature of our model, we can also consider a monodisperse model to get a minimal model. In this case, only four geometric quantities are required, two related to statics and two related to dynamics. Compatibility with Section \ref{sec:two-scale} leads us to consider $(\Sigma\left<G\right>, \alpha_1^d, \DSigPerp, \DtSigPerp)$ or equivalently $(n_1^d, \alpha_1^d, \DSigPerp, \DtSigPerp)$. We then obtain the following quadrature
\begin{equation}
    \label{eq:monodisperse_quad}
    \begin{gathered}
        n(\boldsymbol{x}, t, \widehat{S}_0, \widehat{\chi}, \widehat{\dot{\chi}}) = n_1^d \delta(\widehat{S}_0 - S_0)\delta(\widehat{\chi}-\chi)\delta(\widehat{\dot{\chi}}-\dot{\chi}),\\
        S_0 = \frac{(6\sqrt{\pi}\alpha_1^d)^{2/3}}{(n_1^d)^{2/3}},
        \quad
        \chi = 2\sqrt{\frac{2}{5}}\frac{\DSigPerp}{(n_1^d)^{1/3}(6\sqrt{\pi}\alpha_1^d)^{2/3}},
        \quad
        \dot{\chi}= 2\sqrt{\frac{2}{5}}\frac{\DtSigPerp}{(n_1^d)^{1/3}(6\sqrt{\pi}\alpha_1^d)^{2/3}}.
    \end{gathered}
\end{equation}
All other geometric quantities can then be reconstructed through moments of $n$. The minimization of the Lagrangian is similar to the polydisperse one, and, without adding dissipative source terms, we obtain the following set of equations
\begin{equation}
    \label{eq:final_dissipative_monodisperse_osc}
    \begin{cases}
        \setlength{\arraycolsep}{0pt}
        \begin{array}{llll}
            \partial_t m_k &+ \bnabla \bcdot (m_k \vel)&=0, &\qquad k=1,2,1^d,\\
            \partial_t n_1^d &+ \bnabla \bcdot (n_1^d \vel)&=0,&\\
            \partial_t (n_1^d(S_0)) &+ \bnabla \bcdot (n_1^d(S_0) \vel)&=0,&\\
            \partial_t (n_1^d\dot{\chi}) &+ \bnabla \bcdot (n_1^d\dot{\chi} \vel) &= -\omega^2n_1^d\chi,\\
            \partial_t (n_1^d\chi) &+ \bnabla \bcdot (n_1^d\chi \vel) &=n_1^d\dot{\chi}, & \\
            \partial_t (\rho \vel)&+ \bnabla \bcdot (\rho \vel \otimes \vel+p \boldsymbol{I})&=0,          &          \\
        \end{array}
    \end{cases}
\end{equation}
with $p:=p_1=p_2$ where $\omega^2 = \tilde{\omega}^2(S_0)^{-3/2}$, $\beta = 40\pi\nu_{vis}/S_0$.
We can then provide the dynamics of other geometric quantities such as the interface area density $\Sigma = M_{1,0,0}^{\xi}+M_{1,2,0}^{\xi}$,
\begin{equation}
        \partial_t \Sigma + \bnabla \bcdot (\Sigma \vel) 
        =\partial_t (n_1^d (S_0)\chi^2) + \bnabla \bcdot (n_1^d (S_0)\chi^2 \vel)
        =n_1^d (S_0)D_t (\chi^2)
        =2 n_1^d (S_0)\chi \dot{\chi}.
\end{equation}
Replacing with geometric variables leads to
\begin{equation}
    \partial_t \Sigma + \bnabla \bcdot (\Sigma \vel) = \frac{16}{5}
    \frac{\DSigPerp\DtSigPerp}{(n_1^d)^{1/3}(6\sqrt{\pi}\alpha_1^d)^{2/3}}.
\end{equation}
Remark that we cannot get rid of the oriented surface area density terms in all terms of the evolution equation of $\Sigma$. Moreover, no parameter of the physics is present.